\documentclass{article}

\usepackage{amssymb}
\usepackage{amsmath}
\usepackage{graphicx}

\usepackage{natbib}
\begin{document}

%\onecolumn
\hspace{5.8cm}
Accepted for publication in Astronomy \& Astrophysics, March 18, 2004
\\
\vspace{1cm}
\begin{center}
{\Large \bf Velocity centroids and the structure of interstellar turbulence:\\
I. Analytical study}
\\
\vspace{0.5cm}

{\large Fran\c{c}ois Levrier}
\\
\vspace{0.2cm}

LERMA / LRA, \'Ecole normale sup\'erieure, 24 rue Lhomond, 75231 Paris Cedex 05, France
\\
\vspace{0.8cm}

%\offprints{F. Levrier, \email{levrier@lra.ens.fr}}
%\date{Received 26 January 2004 / Accepted 15 March 2004}

{\bf ABSTRACT}\\
\end{center}
We present an analytical study of the statistical properties of integrated emission and velocity centroids for a slightly compressible turbulent slab model, to retrieve the underlying statistics of three-dimensional density and velocity fluctuations. Under the assumptions that the density and velocity fields are homogeneous and isotropic, we derive the expressions of the antenna temperature for an optically thin spectral line observation, and of its successive moments with respect to the line of sight velocity component, focusing on the zeroth (intensity or integrated emission $I$) and first (non-normalized velocity centroid $C$) moments. The ratio of the latter to the former is the normalized centroid $C_0$, whose expression can be linearized for small density fluctuations. To describe the statistics of $I$, $C$ and $C_0$, we derive expansions of their autocorrelation functions in powers of density fluctuations and perform a lowest-order real-space calculation of their scaling behaviour, assuming that the density and velocity fields are fractional Brownian motions. We hence confirm, within the scope of this study, the property recently found numerically by~\citet{mamd2003} that the spectral index of the normalized centroid is equal to that of the full velocity field. However, it is also argued that, in order to retrieve the velocity statistics, normalization of centroids may actually not be the best way to remove the influence of density fluctuations. In this respect, we discuss the modified velocity centroids introduced by~\citet{lazarian2003} as a possible alternative. In a following paper, we shall present numerical studies aimed at assessing the validity domain of these results. Appendices~\ref{sec_chandra} to~\ref{sec_uniform} are only available online at EdP Sciences.
\\

{\it Subject headings:} ISM: structure -- Methods: analytical -- Turbulence

%\maketitle

\section{Introduction}

The proper exploitation of astronomical observations requires one to deal with several problems to accurately describe the objects and processes under study. In particular, regarding the physics of the interstellar medium (ISM), it should be stressed that spectral emission data depends on the velocity field solely via its component along the line of sight, through Doppler shifts. Furthermore, this information is necessarily integrated along the line of sight, and radiative transfer leads to expressions in which the contributions of the density and velocity fields are mixed in a complex way~\citep{hegmann2000}. Hence, to describe the physical conditions and processes in the ISM, one has to rely on a single number (e.g. antenna temperature) for any given direction in the plane of the sky and any given velocity along the line of sight. Although the comparison of antenna temperatures for various tracers helps, it is necessary to solve an inverse problem to have access to the three-dimensional properties of the medium, such as density and velocity, and compare them with various models, for instance to assess the roles of gravity, magnetic field or turbulence. 

Indeed, it is now recognized that the different components of the ISM are subjected to turbulent motion. This has been observed in the ionized gas~\citep{vanlangevelde92}, in {\sc Hii} regions~\citep{odell87} and in the neutral atomic phase~\citep{spicker88,miville2003}, but molecular cloud studies are by far the most numerous~\citep[see e.g.][]{kleiner85a,kitamura93,miesch94}. Estimates of Reynolds numbers from molecular viscosity in these clouds are of the order of $10^8$, consistent with turbulent flows~\citep{chandrasekhar49}. Moreover, molecular lines in this phase exhibit suprathermal widths over a wide range of spatial scales, from a few km~s$^{-1}$ in small dark clouds to a few tens of km~s$^{-1}$ in giant complexes, while the thermal dispersion is only of about 0.3~km~s$^{-1}$ for molecular hydrogen at $T=10~$K. These linewidths scale as a power-law of the cloud's size~\citep{larson81}, with an exponent close to that predicted by the classical theory of turbulence~\citep{k41}. Such random motions within molecular clouds may in turn account for a number of other properties of spectral lines~\citep{baker76}, as well as provide support against gravitational collapse, explaining the fact that the lifetime of molecular clouds is larger than their free-fall time~\citep{scalo85} and that the star formation rate is therefore much smaller than predicted by gravitationally collapsing cloud models~\citep{zuckerman74}.

Because of this interplay between random motion and many physical processes at work in the ISM in general and in the molecular phase in particular, one needs to accurately describe these turbulent flows~\citep[see the review by][]{vazquez2000}. As noted earlier, this has to be done using the antenna temperature which represents the emission from a given direction and at a given velocity, and so a number of methods were devised to derive the statistical properties of the three-dimensional fields from those of the observational data. Among these, the velocity channel analysis (VCA) of~\citet{lp00} is based on an analytical derivation of the properties of channel maps with varying velocity widths. It may however prove difficult to apply to actual observations, as shown by~\citet{mamd2003}. The modified velocity centroids (MVC) of~\citet{lazarian2003} are a recent promising attempt to reduce the influence of density fluctuations in the statistical properties of centroids, although it can be argued that they are only defined through their structure function. As a final example of velocity statistics retrieval methods, principal component analysis (PCA), which works on the full position-position-velocity cubes, is meant to decompose data onto an orthogonal basis and derive properties of the velocity field at each scale, as calibrated numerically by~\citet{bruntetal2003}. The main objective of these works is to relate the scaling behaviour observed in the two-dimensional maps to scaling laws inferred for the three-dimensional fields. For instance,~\citet{stutzki98} showed that for optically thin media, the spectral index of the integrated emission map is the same as that of the full density field, provided that the depth probed is larger than the transverse scales considered. In a recent work,~\citet{mamd2003} used numerical simulations to show that the same is true for the normalized velocity centroid with respect to the three-dimensional velocity field. However, this latter result lacks theoretical support, and it is therefore the goal of this paper to present an analytical study aimed at clarifying the relationship between the velocity centroids and the velocity field, within a simple turbulent cloud model.

Given the observational data, one may derive moments of the antenna temperature profiles, each of these moments yielding a potentially informative two-dimensional map. For instance, the zeroth moment is the integrated emission, or intensity, while the first moment is the velocity centroid, which can also be normalized to the intensity~\citep{munch58,kleiner85b,mamd2003}. For an optically thin line and uniform excitation conditions, the non-normalized centroid can be related to the cloud's total momentum, while the normalized centroid is a measure of average velocity within the medium. While density fluctuations may bias the description of the turbulent motion~\citep{lazarian2003}, it is however commonly believed, and intuitively plausible, that their effects are somewhat compensated by normalization. To properly assess these, and following~\citet{kleiner85b}~\citep[see also][]{scalo84,kitamura93,miesch94}, we shall use autocorrelation functions of the moment maps and relate them to correlation functions of the underlying three-dimensional fields. To this end, we first describe the model we shall use and introduce the notations and assumptions in section~\ref{sec_model}. We then present a brief summary of how moments of the line profile can be related to the density and velocity fields within the medium (section~\ref{sec_moments}). Section~\ref{sec_norm} contains a general study of the statistical properties of intensity, normalized and non-normalized velocity centroid maps as functions of the three-dimensional density and velocity fields' statistics. The equations obtained in the lowest order are then applied to the test case of fractional Brownian motion (fBm) density and velocity fields (section~\ref{sec_fbm}). Section~\ref{sec_test} presents a discussion of the various results obtained with respect to earlier works. Our concluding remarks are given in section~\ref{sec_concl} and details on the calculations can be found in the appendix.

\begin{center}
\begin{figure}[htbp]
\begin{picture}(16,7)
\put(0.8,0){\includegraphics[width=16cm]{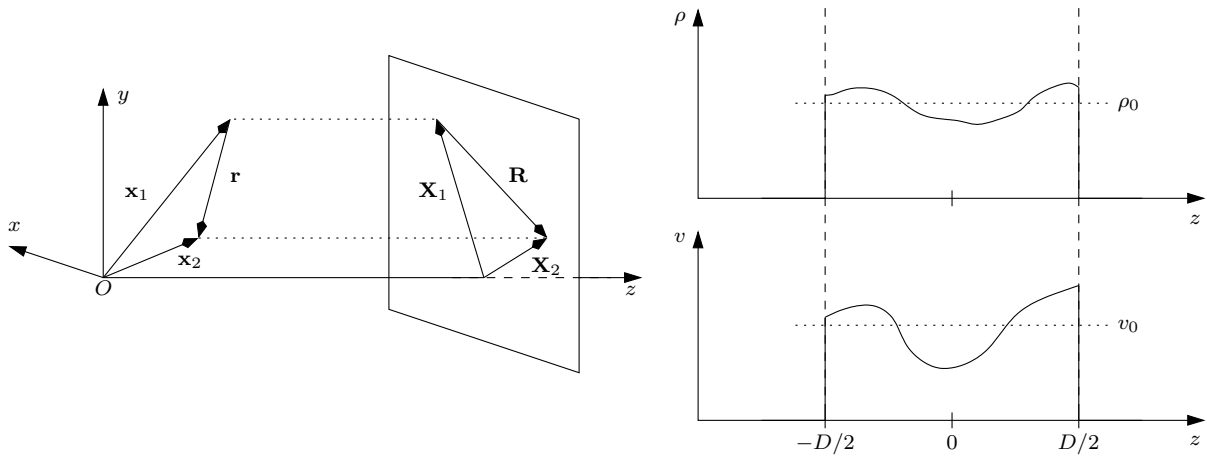}}
\put(0.85,2.95){\footnotesize{$x$}}
\put(2.3,4.7){\footnotesize{$y$}}
\put(9.05,2.1){\footnotesize{$z$}}
\put(2,2.1){\footnotesize{$O$}}
\put(3.8,3.6){\footnotesize{$\mathbf{r}$}}
\put(2.4,3.4){\footnotesize{$\mathbf{x}_1$}}
\put(3.1,2.5){\footnotesize{$\mathbf{x}_2$}}
\put(7.5,3.6){\footnotesize{$\mathbf{R}$}}
\put(6.3,3.4){\footnotesize{$\mathbf{X}_1$}}
\put(7.8,2.4){\footnotesize{$\mathbf{X}_2$}}
\put(16.55,0.1){\footnotesize{$z$}}
\put(16.55,3.05){\footnotesize{$z$}}
\put(13.32,0.05){\footnotesize{$0$}}
\put(11.32,0.05){\footnotesize{$-D/2$}}
\put(14.8,0.05){\footnotesize{$D/2$}}
\put(9.7,5.75){\footnotesize{$\rho$}}
\put(9.7,2.8){\footnotesize{$v$}}
\put(15.6,1.65){\footnotesize{$v_0$}}
\put(15.6,4.6){\footnotesize{$\rho_0$}}
\end{picture}
\caption{\label{fig_1} Notations used in the paper (left) and schematics of the turbulent slab model (right). A lowercase boldface letter stands for a three-dimensional vector, while the corresponding uppercase represents its projection on the plane of the sky ($xOy$). The slab is infinite in the $x$ and $y$ directions and is limited to $I_D=[-D/2,D/2]$ in the $z$ direction. The density $\rho$ and line-of-sight velocity $v$ distributions along a line of sight are shown on the right. The mean values $\rho_0$ and $v_0$ are taken over the whole slab. }
\end{figure}   
\end{center}

\section{The model}
\label{sec_model}

Throughout the paper, boldface notations stand for vector quantities. A point in three-dimensional space is given by its position $\mathbf{x}$, which can be written as $\mathbf{x}=(\mathbf{X},z)=\mathbf{X}+z\mathbf{e}_z$, where $\mathbf{X}$ is a two-dimensional vector in the plane of the sky and $z$ marks the line-of-sight position of the point considered, $\mathbf{e}_z$ being the unit vector along the line of sight. The three-dimensional separation between points $\mathbf{x}_1$ and $\mathbf{x}_2$ is written as $\mathbf{r}=\mathbf{x}_2-\mathbf{x}_1$ and the separation between their respective lines of sight $\mathbf{X}_1$ and $\mathbf{X}_2$ is $\mathbf{R}=\mathbf{X}_2-\mathbf{X}_1$. In short, three-dimensional vectors are written in lowercase, while vectors in the plane of the sky are written in uppercase. These notations are illustrated in Fig~\ref{fig_1}. It should be emphasized that we shall only consider small scales on the plane of the sky, so that all lines of sight are parallel. 

Hereafter, our model is a slab of gas of width $D$, perpendicular to the line of sight, and of infinite transverse extensions\footnote{This is in no way contradictory with the assumption that all lines of sight considered are parallel. The cloud is supposed to be infinite, but only a small fraction of it is observed and supposed to be representative of the whole.}, with $z=0$ conventionally placed halfway through the slab (see Fig.~\ref{fig_1}). Within the slab, the average density (that is, the average number of emitters per unit volume taken over the whole slab) is a constant noted $\rho_0$. As for the velocity $\mathbf{v}$ of the gas with respect to the observer, we shall write $v$ to stand for its component along the line of sight, the mean value of which over the slab is $v_0$. These averages ($\rho_0$ and $v_0$) are obviously constants, independent of the line of sight. To assume the most general case, we shall take $v_0$ to be nonzero\footnote{The case $v_0=0$ is only special in the way the calculations are performed, as the end result amounts to setting $v_0$ to zero in the general formul\ae. This is shown in appendix~\ref{sec_zero}.}. The gas velocity within the rest frame of the cloud is simply $\mathbf{\delta v}=\mathbf{v}-v_0\mathbf{e}_z$, assuming that there is no systematic transverse velocity. Furthermore, fluctuations of density along any given line of sight are supposed to be small compared to the average value $\rho_0$, in the sense that the variance of the density field should be less than $\rho_0$. It is also assumed that the turbulent flow within the slab is homogeneous and isotropic. These last hypotheses are to be understood in the strong sense of~\citet{monin75}: not only are the scalar field $\rho$ and the vector field $\mathbf{\delta v}$ homogeneous and isotropic, but the same should apply to the four-dimensional field $(\rho,\mathbf{\delta v})$. This will be useful in the derivation of our results in section~\ref{sec_norm}. Outside the slab, the density and velocity are set to zero. We further assume that the observation is done in an optically thin transition, at a frequency for which the Rayleigh-Jeans approximation is valid, that uniform excitation conditions apply within the slab, and that no background radiation is present.

With these assumptions in mind, the antenna temperature $T_\mathrm{a}(\mathbf{X},u)$ representing the emission from a given line of sight $\mathbf{X}$ at an observed velocity $u$ can be written as the integral~\citep{kleiner85b},
\begin{equation}
\label{eq_4b}
T_\mathrm{a}(\mathbf{X},u)=\int\limits_{I_D} \! T_\mathrm{ex}(\mathbf{x}) \kappa_0(\mathbf{x})\phi(v(\mathbf{x})-u)\mathrm{d} z,
\end{equation}
where $I_D$ is the segment $[-D/2,D/2]$ over which the integration is performed. At each position $\mathbf{x}$, $T_\mathrm{ex}$ is the excitation temperature and $\kappa_0$ is the integrated absorption coefficient. The normalized line profile function $\phi$ is assumed to be symmetrical about zero and independent of $\mathbf{x}$. For instance, if one only considers thermal broadening, $\phi$ takes the form
\begin{equation}
\phi(w)=\frac{1}{\sqrt{2\pi}\sigma_\mathrm{th}}\exp{\left[-\frac{w^2}{2\sigma_\mathrm{th}^2}\right]},
\end{equation}
with $\sigma_\mathrm{th}$ the thermal velocity dispersion. The shifted argument $v(\mathbf{x})-u$ of $\phi$ in Eq.~(\ref{eq_4b}) simply expresses the fact that the emission at position $\mathbf{x}$ is broadened around the local line of sight velocity  $v(\mathbf{x})$. Assuming uniform excitation conditions, $T_\mathrm{ex}$ is a constant, and $\kappa_0(\mathbf{x})$ is proportional to the local gas density $\rho(\mathbf{x})$, so we can write
\begin{equation}
\label{eq_4}
T_\mathrm{a}(\mathbf{X},u)=\int\limits_{I_D}  \!\alpha\rho(\mathbf{x})\phi(v(\mathbf{x})-u)\mathrm{d} z,
\end{equation}
where $\alpha$ is a proportionality constant. The expression in Eq.~(\ref{eq_4}) shows that, even under simplifying assumptions, the task of extracting properties of the density and velocity fields from the observational data $T_\mathrm{a}(\mathbf{X},u)$ is a difficult one. 

\section{Velocity moments}
\label{sec_moments}

In order to obtain information about the velocity field properties from the data sets, one may compute the various moments of the antenna temperature profiles $T_\mathrm{a}(\mathbf{X},u)$ for a given line of sight $\mathbf{X}$. This is a logical step considering that knowledge of the moments of a random variable is equivalent to that of the full probability distribution. We shall therefore consider the moment maps $W_n(\mathbf{X})$ defined by
\begin{equation}
W_n(\mathbf{X})=\int \!\! u^n T_\mathrm{a}(\mathbf{X},u) \mathrm{d} u.
\end{equation}
The integration is done over all velocities from $-\infty$ to $\infty$, which poses no convergence problem since the line profile has a finite support\footnote{If any continuum emission is present, it should first be removed.}. Using Eq.~(\ref{eq_4}) to express $T_\mathrm{a}$ as a function of the local line profile and of the density,
\begin{equation}
W_n(\mathbf{X})=\int \!\! u^n \!\! \left[ \alpha \!\!\int\limits_{I_D} \!\!\! \rho(\mathbf{x})\phi(v(\mathbf{x})-u)\mathrm{d} z \right] \mathrm{d} u=\alpha \!\!\int\limits_{I_D} \!\!\! \rho(\mathbf{x}) \left[\int\limits_{\phantom{I_D}} \!\!\!  \left(v(\mathbf{x})-w\right)^n \phi(w) \mathrm{d} w\right] \mathrm{d} z,
\end{equation}
where we introduced the variable $w=v(\mathbf{x})-u$. Developing the integrand in the innermost integral, and since the local line profile $\phi$ is assumed to be independent of the position $\mathbf{x}$, we get the following expression
 \begin{equation}
W_n(\mathbf{X})=\alpha \sum_{k=0}^n (-1)^kC_n^k \left[\;\, \int\limits_{I_D} \!\!\! \rho(\mathbf{x})v^{n-k}(\mathbf{x})\mathrm{d} z \right]\left[\int\limits_{\phantom{I_D}} \!\!\! w^k \phi(w) \mathrm{d} w\right]=\sum_{k=0}^n C_n^k h_k \left[\;\int\limits_{I_D} \!\!\!  \rho(\mathbf{x})v^{n-k}(\mathbf{x})\mathrm{d} z \right],
\end{equation}
where the $C_n^k$ are the binomial's coefficients and the $h_k$ are related to the moments of the normalized line profile $\phi$,
\begin{equation}
\label{eq_6}
h_k=(-1)^k\alpha\!\!\int \!\! w^k \phi(w) \mathrm{d} w.
\end{equation}
It should be noted that the assumed evenness of $\phi$ leads to $h_{2p+1}=0$. The first two moments are of particuler interest, since, by definition, they are the total emergent intensity $I$ and the non-normalized velocity centroid $C$, respectively,
\begin{equation}
\label{eq_60}
I(\mathbf{X})=W_0(\mathbf{X})=  \alpha \!\!\int\limits_{I_D} \!\!\! \rho(\mathbf{x})\mathrm{d} z=\alpha N(\mathbf{X}) \qquad \mathrm{and} \qquad
C(\mathbf{X})=W_1(\mathbf{X})=  \alpha \!\!\int\limits_{I_D} \!\!\! \rho(\mathbf{x})v(\mathbf{x})\mathrm{d} z.
\end{equation}
Here, $N(\mathbf{X})$ is the column density at position $\mathbf{X}$. In the rest of this paper, we shall only consider the intensity and the velocity centroid. The use of higher order moments would theoretically give access to more information about the structures of the density and velocity fields, but, in practice, noise levels and limited spectral resolution may very well jeopardize their usefulness.

\section{Statistics of intensity and centroid maps}
\label{sec_norm}
\subsection{Rationale of the computations}

We may use the line profile moments defined above to quantify the structure of the turbulent flow within the slab. Obviously, the zeroth moment $W_0(\mathbf{X)}=I(\mathbf{X)}$ can only be used to derive statistics of the density field, as it is proportional to the column density $N(\mathbf{X)}$. The first moment or non-normalized velocity centroid $W_1(\mathbf{X)}=C(\mathbf{X)}$ is the first appropriate quantity to study the velocity field, as can be seen from Eq.~(\ref{eq_60}). However, $C(\mathbf{X)}$ represents the integration of a momentum, rather than of a velocity proper, and it appears that density fluctuations may affect the estimation of the velocity statistics from this map. To circumvent this problem, it is common to use normalized centroids $C_0(\mathbf{X)}$, which are simply defined as the ratio $C(\mathbf{X)}/I(\mathbf{X)}$. This is usually and empirically justified by the assumption that density fluctuations in $C(\mathbf{X)}$ are also present in $I(\mathbf{X)}$ and therefore somehow vanish from $C_0(\mathbf{X)}$. In the simplified case where density does not fluctuate along the line of sight, such a reasoning is obviously correct, even if transverse density fluctuations are present, and normalization then only serves as a dimensionality factor. As far as we are aware, however, no analytical study exists of the influence of longitudinal fluctuations of density on velocity centroids, even in the quite simplified model used here. 

Now, in order to gain a better understanding of the underlying three-dimensional statistics of the velocity field through velocity centroids, it seems natural to consider the two-dimensional statistics of the centroid maps. Indeed, structural properties of the velocity and density fields should arise, under one form or another, in statistical measures performed on the maps $C(\mathbf{X)}$ and $C_0(\mathbf{X)}$. Of course, it is similarly reasonable to look for the influence of density structure in the statistics of the intensity map $I(\mathbf{X)}$. One such useful measure to be performed on the available two-dimensional maps is the autocorrelation function (ACF), which gives the mean degree of correlation between values of a field taken at points separated on the plane of the sky by a given vector $\mathbf{R}$~\citep{kleiner84}. More precisely, the autocorrelation function $A_\mathcal{F}$ of a field $\mathcal{F}$ is defined by
\begin{equation}
A_{\mathcal{F}}(\mathbf{R})=\left< \mathcal{F}(\mathbf{X})\mathcal{F}(\mathbf{X}+\mathbf{R})\right>,
\end{equation}
where the brackets stand for a spatial average over position $\mathbf{X}$ in the plane of the sky. Hereafter, we shall consider the autocorrelation functions of the intensity and of both types of centroid, respectively noted $A_{I}(\mathbf{R})$, $A_{C}(\mathbf{R})$ and $A_{C_0}(\mathbf{R})$, and compute them as functions of statistical measures of the three-dimensional density and velocity fields. In order to do so, we first separate mean and fluctuating contributions of density and velocity in the quantities involved, with $\rho=\rho_0+\delta \rho$ and $v=v_0+\delta v$. These expressions can be used to write the first two moment maps as
\begin{equation}
\label{eq_000}
\!\!\!\!I(\mathbf{X})=\int\limits_{I_D} \!\! \alpha \rho(\mathbf{X},z)\mathrm{d} z=\alpha \rho_0  D \left[1+y_{\rho_\mathbf{X}}\right] \quad \mathrm{and} \quad C(\mathbf{X})=\int\limits_{I_D} \!\! \alpha \rho(\mathbf{X},z)v(\mathbf{X},z)\mathrm{d} z=\alpha \rho_0 v_0 D \left[1+y_{\rho_\mathbf{X}}+y_{v_\mathbf{X}}+y_{\rho v_\mathbf{X}}\right],
\end{equation}
where we have introduced the following integrated fluctuation terms\footnote{In these integrated fluctuation terms, we chose to denote the position $\mathbf{X}$ on the plane of the sky as a subindex, rather than as an argument, to allow for concise equations in the subsequent developments.}
\begin{equation}
y_{\rho_\mathbf{X}}=\frac{1}{D}\int\limits_{I_D} \!\frac{\delta \rho(\mathbf{X},z)}{\rho_0}\mathrm{d} z \; , \quad  y_{v_\mathbf{X}}=\frac{1}{D}\int\limits_{I_D} \!\frac{\delta v(\mathbf{X},z)}{v_0}\mathrm{d} z \quad \mathrm{and} \quad y_{\rho v_\mathbf{X}}= \frac{1}{D}\int\limits_{I_D} \!\frac{\delta \rho(\mathbf{X},z)}{\rho_0}\frac{\delta v(\mathbf{X},z)}{v_0}\mathrm{d} z.
\end{equation}
The normalized centroid $C_0$ is then simply written as
\begin{equation}
\label{eq_301}
C_0(\mathbf{X})=\frac{C(\mathbf{X})}{I(\mathbf{X})}=v_0\frac{1+y_{\rho_\mathbf{X}}+y_{v_\mathbf{X}}+y_{\rho v_\mathbf{X}}}{1+y_{\rho_\mathbf{X}}}.
\end{equation}
This last expression can be used to clarify the usefulness of the small fluctuations hypothesis. Indeed, for a perturbative method to be applicable in the computation of the autocorrelation function of the normalized velocity centroid map, one should be able to linearize Eq.~(\ref{eq_301}), and therefore we need to assume that $|y_{\rho_\mathbf{X}}| < 1$, so that the denominator can be expanded as a converging series,
\begin{equation}
\label{eq_201}
\frac{1}{1+y_{\rho_\mathbf{X}}}=1-y_{\rho_\mathbf{X}}+y_{\rho_\mathbf{X}}^2 \ldots = \sum_{n \geqslant 0} (-y_{\rho_\mathbf{X}})^n.
\end{equation}
Such a condition is obviously achieved if local density fluctuations themselves are small in the sense, expressed in section~\ref{sec_model}, that the standard deviation of the density field $\sigma_\rho$ should be smaller than the average density $\rho_0$, since
\begin{equation}
|y_{\rho_\mathbf{X}}|=\left|\frac{1}{D}\int\limits_{I_D} \frac{\delta \rho(\mathbf{X},z)}{\rho_0}\mathrm{d} z \right| \leqslant \sqrt{\frac{1}{D}\int\limits_{I_D}\!\! \left(\frac{\delta \rho(\mathbf{X},z)}{\rho_0}\right)^2\!\!\mathrm{d} z} \simeq \sqrt{\frac{\overline{\delta\rho^2}}{\rho_0^2}}=\frac{\sigma_\rho}{\rho_0}.
\end{equation}
The identification of the mean squared density fluctuations along a single line of sight with the variance $\sigma_\rho^2$ of the whole density field is based on the homogeneity hypothesis. Indeed, any line of sight should statistically represent the full field. Although the stronger assumption $\sigma_\rho < \rho_0$ is not strictly necessary for the linearization mentioned earlier, it is useful in the expansion of the autocorrelation functions since, in this case, the hybrid integrated fluctuation term $y_{\rho v_\mathbf{X}}$ is of the same order in density as $y_{\rho_\mathbf{X}}$,
\begin{equation}
|y_{\rho v_\mathbf{X}}| \leqslant \sqrt{\frac{1}{D}\int\limits_{I_D}\!\! \left(\frac{\delta \rho(\mathbf{X},z)}{\rho_0}\right)^2\!\!\mathrm{d} z}\sqrt{\frac{1}{D}\int\limits_{I_D}\!\! \left(\frac{\delta v(\mathbf{X},z)}{v_0}\right)^2\!\!\mathrm{d} z} \simeq \frac{\sigma_\rho}{\rho_0}\frac{\sigma_v}{v_0},
\end{equation}
where we used the Cauchy-Schwarz inequality and introduced the standard deviation $\sigma_v$ of the line-of-sight component of the velocity field\footnote{It should be stressed that no hypothesis is made on the strength of velocity fluctuations, so that scaling them to the mean velocity $v_0$ is merely a convenient way to symmetrize expressions, and should not be given too much importance. We are aware, however, that the sound speed would be a much more physically meaningful velocity scale, and it will be used as such when the mean velocity is zero (see appendix~\ref{sec_zero}).}. This allows us to develop the intensity and the velocity centroid maps, as well as their autocorrelation functions, according to the powers of $\sigma_\rho/\rho_0$. It should be made clear that the expansion in Eq.~(\ref{eq_201}) is not necessarily true for real data, but that it is used here as a reasonable first step towards understanding the effects of realistic density fluctuations on velocity centroids.

\begin{center}
\begin{figure}[htbp]
\begin{picture}(16,7)
\put(0.8,0){\includegraphics[width=16cm]{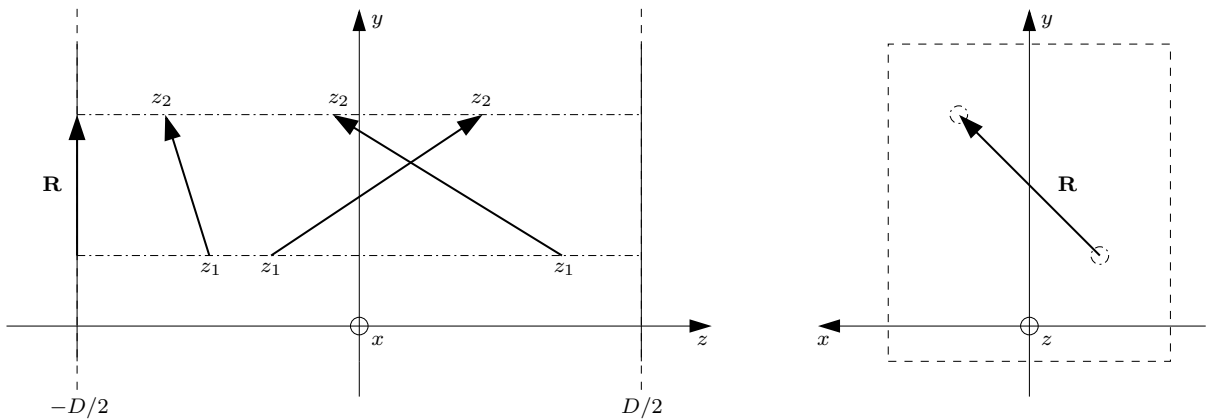}}
\put(5.68,0.95){\footnotesize{$x$}}
\put(5.68,5.22){\footnotesize{$y$}}
\put(10,0.95){\footnotesize{$z$}}
\put(2.75,4.17){\footnotesize{$z_2$}}
\put(5.1,4.17){\footnotesize{$z_2$}}
\put(7,4.17){\footnotesize{$z_2$}}
\put(3.4,1.93){\footnotesize{$z_1$}}
\put(4.2,1.93){\footnotesize{$z_1$}}
\put(8.1,1.93){\footnotesize{$z_1$}}
\put(1.3,3){\footnotesize{$\mathbf{R}$}}
\put(1.4,0.05){\footnotesize{$-D/2$}}
\put(9,0.05){\footnotesize{$D/2$}}
\put(14.58,0.95){\footnotesize{$z$}}
\put(11.6,0.95){\footnotesize{$x$}}
\put(14.58,5.22){\footnotesize{$y$}}
\put(14.8,3){\footnotesize{$\mathbf{R}$}}
\end{picture}
\caption{\label{fig_2} Interpretation of the average $M_{\rho,\rho}$ appearing in Eq.~(\ref{eq_24}). The slab is viewed edge-on in the left figure, and face-on in the right. The value of $M_{\rho,\rho}$ at any lag $\mathbf{R}$ is the average of the correlation function $B_{\rho,\rho}$ over all pairs of points whose projected separation on the plane of the sky is $\mathbf{R}$. Three such pairs are presented on the left.}
\end{figure}   
\end{center}

\subsection{Autocorrelation function of the intensity $I$}

Taking the autocorrelation function $A_I(\mathbf{R})$ of the intensity map $I(\mathbf{X})$, and using the linearity property of averages, we have terms of order up to two in density fluctuations, namely
\begin{equation}
A_I(\mathbf{R})=\left<I(\mathbf{X})I(\mathbf{X}+\mathbf{R})\right>=(\alpha\rho_0D)^2\left[1+\left<y_{\rho_{\mathbf{X}\phantom{+}}}\!\!\!\right>+\left<y_{\rho_{\mathbf{X}+\mathbf{R}}}\right>+\left<y_{\rho_\mathbf{X}} y_{\rho_{\mathbf{X}+\mathbf{R}}}\right>\right],
\end{equation}
where we recall that the brackets stand for an average over $\mathbf{X}$. These are performed on quantities integrated over $z$, and can therefore be interpreted as averages over the whole turbulent slab. Indeed, if one considers a three-dimensional field $f(\mathbf{x})$, its integrated map $F(\mathbf{X})$ defined by
\begin{equation}
\label{eq_1}
F(\mathbf{X})=\frac{1}{D}\int\limits_{I_D} \! f(\mathbf{X},z)\mathrm{d} z \quad \mathrm{has~an~average} \quad \left<F(\mathbf{X})\right>=\frac{1}{D}\int\limits_{I_D} \! \left<f(\mathbf{X},z)\right>\mathrm{d} z=\frac{1}{DS}\int\!\!\!\!\int\!\!\!\!\int\! f(\mathbf{x})\mathrm{d} \mathbf{x}
\end{equation}
over a surface $S$ of the sky. The quantity $\left<F(\mathbf{X})\right>$ can then be seen as an average of $f$ over the volume $DS$, and, assuming ergodicity, it can also be identifid with the ensemble average $\overline{f(\mathbf{x})}$ at any given position $\mathbf{x}$ within the flow. In the present case, the term $\left<y_{\rho_{\mathbf{X}\phantom{+}}}\!\!\!\right>$, which is of the first order in density fluctuations, can be written as
\begin{equation}
\left<y_{\rho_{\mathbf{X}\phantom{+}}}\!\!\!\right>=\frac{1}{\rho_0 D}\int\limits_{I_D} \! \left<\delta \rho(\mathbf{X},z)\right>\mathrm{d} z=\frac{1}{\rho_0}\overline{\delta \rho(\mathbf{x})}=0,
\end{equation}
since, by definition, the average of the density fluctuations over the whole volume of the slab is zero. The assumption of homogeneity allows one to write that $\left<y_{\rho_{\mathbf{X}+\mathbf{R}}}\right>=0$ as well. As for the second order term, which reads
\begin{equation}
\label{eq_9}
\left<y_{\rho_\mathbf{X}} y_{\rho_{\mathbf{X}+\mathbf{R}}}\right>=\frac{1}{\rho_0^2 D^2}\left<\left[\;\int\limits_{I_D} \! \delta \rho(\mathbf{X},z)\mathrm{d} z\right]\left[\;\int\limits_{I_D} \! \delta \rho({\mathbf{X}+\mathbf{R}},z)\mathrm{d} z\right] \right>=\frac{1}{\rho_0^2 D^2}\int\!\!\!\!\int\limits_{\!\!I^2_D} \! \left<\delta \rho(\mathbf{X},z_1) \delta \rho(\mathbf{X}+\mathbf{R},z_2)\right> \mathrm{d} z_1 \mathrm{d} z_2,
\end{equation}
and where $I^2_D$ is the square domain $I_D \times I_D$, it includes an average over $\mathbf{X}$ that, following the idea used above, should be replaced by an ensemble average characteristic of the turbulent flow. Indeed, one can see that
\begin{equation}
\left<\delta \rho(\mathbf{X},z_1) \delta \rho(\mathbf{X}+\mathbf{R},z_2)\right>=\overline{\delta \rho(\mathbf{x}_1) \delta \rho(\mathbf{x}_2)},
\end{equation}
with the three-dimensional vectors $\mathbf{x}_1=(\mathbf{X},z_1)$ and $\mathbf{x}_2=(\mathbf{X}+\mathbf{R},z_2)$. Introducing the autocorrelation function $B_{\rho,\rho}$ of the density fluctuations, which is defined, for any pair of points $(\mathbf{x},\mathbf{x}+\mathbf{r})$ within the slab, by\footnote{We chose to write $B_{\rho,\rho}$ instead of using the more consistent notation $A_{\delta\rho}$ for the autocorrelation function of the density fluctuations, because, in the following developments, higher order and multipoint correlation functions of mixed fields arise for which the $A_\mathcal{F}$ notation would have become cumbersome.}
\begin{equation}
B_{\rho,\rho}(\mathbf{r})=\overline{ \delta \rho(\mathbf{x}) \delta \rho(\mathbf{x}+\mathbf{r})},
\end{equation}
we therefore can write the term under study as
\begin{equation}
\label{eq_10}
\left<y_{\rho_\mathbf{X}} y_{\rho_{\mathbf{X}+\mathbf{R}}}\right>=\frac{1}{\rho_0^2 D^2}\int\!\!\!\!\int\limits_{\!\!I^2_D} \! B_{\rho,\rho}(\mathbf{x}_2-\mathbf{x}_1)\mathrm{d} z_1 \mathrm{d} z_2 \quad \mathrm{or,~more~concisely,} \quad \left<y_{\rho_\mathbf{X}} y_{\rho_{\mathbf{X}+\mathbf{R}}}\right>=\frac{1}{\rho_0^2} M_{\rho,\rho}(\mathbf{R}),
\end{equation}
where $M_{\rho,\rho}(\mathbf{R})$, which is defined by the double integral
\begin{equation}
\label{eq_24}
M_{\rho,\rho} (\mathbf{R})=\frac{1}{D^2}\int\!\!\!\!\int\limits_{\!\!I^2_D} \! B_{\rho,\rho}(\mathbf{R}+(z_2-z_1)\mathbf{e}_z)\mathrm{d} z_1 \mathrm{d} z_2,
\end{equation}
is interpreted as an average of $B_{\rho,\rho}(\mathbf{r})$ taken over all pairs of points within the slab whose three-dimensional separation $\mathbf{r}$ has the given vector $\mathbf{R}$ for component in the plane of the sky (see Fig.~\ref{fig_2}). Eventually, the autocorrelation function of the intensity simply reads
\begin{equation}
\label{eq_002}
A_I(\mathbf{R})=(\alpha D)^2\left[\rho_0^2+M_{\rho,\rho}(\mathbf{R})\right].
\end{equation}
This expression will be exploited later on when compared with the autocorrelation functions of velocity centroids.

\subsection{Autocorrelation function of the non-normalized velocity centroid $C$}

The same general method applies when considering the non-normalized velocity centroid $C(\mathbf{X})$. Its expansion, written in Eq.~(\ref{eq_000}), includes terms of order zero and one in density fluctuations. Therefore, its autocorrelation function $A_C(\mathbf{R})$ features terms of order up to two. Namely,
\begin{equation}
A_C(\mathbf{R})=(\alpha \rho_0 v_0 D)^2\left<\left[1+y_{v_\mathbf{X}}+y_{\rho_\mathbf{X}}+y_{\rho v_\mathbf{X}}\right]\left[1+y_{v_{\mathbf{X}+\mathbf{R}}} +y_{\rho_{\mathbf{X}+\mathbf{R}}}+y_{\rho v_{\mathbf{X}+\mathbf{R}}}\right]\right>=(\alpha \rho_0 v_0 D)^2\sum_{n=0}^2\left<a_{n}\right>,
\end{equation}
where $a_n$ is a term of order $n$ in density. The explicit forms of these coefficients are given by
\begin{equation}
\label{eq_11}
\begin{array}{l}
a_{0}= 1+y_{v_\mathbf{X}}+y_{v_{\mathbf{X}+\mathbf{R}}}+y_{v_\mathbf{X}} y_{v_{\mathbf{X}+\mathbf{R}}}, \\
a_{1}=y_{\rho_\mathbf{X}}+y_{\rho_{\mathbf{X}+\mathbf{R}}}+y_{\rho_\mathbf{X}} y_{v_{\mathbf{X}+\mathbf{R}}}+y_{v_\mathbf{X}}y_{\rho_{\mathbf{X}+\mathbf{R}}}+y_{\rho v_\mathbf{X}}+y_{\rho v_{\mathbf{X}+\mathbf{R}}}+y_{\rho v_\mathbf{X}} y_{v_{\mathbf{X}+\mathbf{R}}}+y_{\rho v_{\mathbf{X}+\mathbf{R}}} y_{v_\mathbf{X}}, \\
a_{2}=y_{\rho_\mathbf{X}} y_{\rho_{\mathbf{X}+\mathbf{R}}}+y_{\rho v_\mathbf{X}} y_{\rho_{\mathbf{X}+\mathbf{R}}}+y_{\rho v_{\mathbf{X}+\mathbf{R}}}y_{\rho_\mathbf{X}}+y_{\rho v_\mathbf{X}} y_{\rho v_{\mathbf{X}+\mathbf{R}}}.
\end{array}
\end{equation}
The linearity of the averaging process over the plane of the sky implies then that we should consider terms of the form $\left<y_{\lambda_{\mathbf{X}\phantom{+}}}\!\!\!\right>$ and $\left<y_{\lambda_\mathbf{X}} y_{\mu_{\mathbf{X}+\mathbf{R}}}\right>$, where $\lambda$ and $\mu$ represent either $\rho$, $v$ or $\rho v$. Due to the homogeneity hypothesis, terms of the form $\left<y_{\lambda_{\mathbf{X}+\mathbf{R}}}\right>$ are of course equal to $\left<y_{\lambda_{\mathbf{X}\phantom{+}}}\!\!\!\right>$, and so, considering the zeroth order contribution $\left<a_0\right>$, we have
\begin{equation}
\left<a_0\right>=1+2\left<y_{v_{\mathbf{X}\phantom{+}}}\!\!\!\right>+\left<y_{v_\mathbf{X}} y_{v_{\mathbf{X}+\mathbf{R}}} \right>=1+\frac{2}{v_0}\overline{\delta v(\mathbf{x})}+\frac{1}{v_0^2}M_{v,v}(\mathbf{R})=1+\frac{1}{v_0^2}M_{v,v}(\mathbf{R}),
\end{equation}
with $M_{v,v}(\mathbf{R})$ being defined similarly to $M_{\rho,\rho}(\mathbf{R})$ in Eq.~(\ref{eq_24}),
\begin{equation}
M_{v,v} (\mathbf{R})=\frac{1}{D^2}\int\!\!\!\!\int\limits_{\!\!I^2_D} \! B_{v,v}(\mathbf{R}+(z_2-z_1)\mathbf{e}_z)\mathrm{d} z_1 \mathrm{d} z_2,
\end{equation}
and $B_{v,v}(\mathbf{r})=\overline{ \delta v(\mathbf{x}) \delta v(\mathbf{x}+\mathbf{r})}$ is the autocorrelation function of the fluctuations of the line-of-sight velocity component. Turning to the first order term, we have, since $\left<y_{\rho_{\mathbf{X}\phantom{+}}}\!\!\!\right>=\left<y_{\rho_{\mathbf{X}+\mathbf{R}}}\right>=0$ as shown in the previous section,
\begin{equation}
\left<a_1\right>=\left<y_{\rho_\mathbf{X}} y_{v_{\mathbf{X}+\mathbf{R}}} \right>+\left<y_{v_\mathbf{X}} y_{\rho_{\mathbf{X}+\mathbf{R}}}\right>+2\left<y_{\rho v_{\mathbf{X}\phantom{+}}}\!\!\!\right>+\left<y_{\rho v_\mathbf{X}} y_{v_{\mathbf{X}+\mathbf{R}}} \right>+\left<y_{\rho v_{\mathbf{X}+\mathbf{R}}} y_{v_\mathbf{X}}\right>
\end{equation}
Each term in this equation can be related to correlation functions of the fluctuation fields $\delta \rho$ and $\delta v$, beginning with  
\begin{equation}
\left<y_{\rho_\mathbf{X}} y_{v_{\mathbf{X}+\mathbf{R}}} \right>=\frac{1}{\rho_0 v_0 D^2}\int\!\!\!\!\int\limits_{\!\!I^2_D} \! \left<\delta \rho(\mathbf{X},z_1) \delta v(\mathbf{X}+\mathbf{R},z_2)\right> \mathrm{d} z_1 \mathrm{d} z_2=\frac{1}{\rho_0 v_0 D^2}\int\!\!\!\!\int\limits_{\!\!I^2_D} \! B_{\rho,v}(\mathbf{R}+(z_2-z_1)\mathbf{e}_z)\mathrm{d} z_1 \mathrm{d} z_2,
\end{equation}
introducing the mixed correlation function $B_{\rho,v}(\mathbf{r})=\overline{ \delta \rho(\mathbf{x}) \delta v(\mathbf{x}+\mathbf{r})}$. Considering the next term $\left<y_{v_\mathbf{X}} y_{\rho_{\mathbf{X}+\mathbf{R}}} \right>$, the homogeneity hypothesis allows us to shift the arguments by $-\mathbf{R}$ so that, exchanging the integration variables,
\begin{equation}
\left<y_{v_\mathbf{X}} y_{\rho_{\mathbf{X}+\mathbf{R}}} \right>=\left<y_{\rho_\mathbf{X}} y_{v_{\mathbf{X}-\mathbf{R}}} \right>=\frac{1}{\rho_0 v_0 D^2}\int\!\!\!\!\int\limits_{\!\!I^2_D} \! B_{\rho,v}(-\mathbf{R}+(z_1-z_2)\mathbf{e}_z)\mathrm{d} z_1 \mathrm{d} z_2,
\end{equation}
Combination of both terms leads to
\begin{equation}
\label{eq_23}
\left<y_{v_\mathbf{X}} y_{\rho_{\mathbf{X}+\mathbf{R}}} \right>+\left<y_{\rho_\mathbf{X}} y_{v_{\mathbf{X}+\mathbf{R}}} \right>=\frac{1}{\rho_0 v_0 D^2}\int\!\!\!\!\int\limits_{\!\!I^2_D} \! \left[B_{\rho,v}(\mathbf{R}+(z_2-z_1)\mathbf{e}_z)+B_{\rho,v}(-\mathbf{R}+(z_1-z_2)\mathbf{e}_z)\right]\mathrm{d} z_1 \mathrm{d} z_2.
\end{equation}
Now, according to~\citet{monin75}, and assuming that the four-dimensional field $(\rho,\mathbf{\delta v})$, made up of the density $\rho$ and vector velocity $\mathbf{\delta v}$, is homogeneous and isotropic, the correlation function of $\rho$ and $\mathbf{\delta v}$ is a vector quantity, and due to isotropy should be along the separation vector $\mathbf{r}$. It should therefore be of the form $\overline{\rho(\mathbf{x}) \mathbf{\delta v}(\mathbf{x}+\mathbf{r})}=f(r)\mathbf{r}$, so that, projecting this relation on $\mathbf{e}_z$, we have
\begin{equation}
f(r)\mathbf{r}.\mathbf{e}_z=\overline{\rho(\mathbf{x}) \mathbf{\delta v}(\mathbf{x}+\mathbf{r})}.\mathbf{e}_z=\overline{\delta\rho(\mathbf{x}) \mathbf{\delta v}(\mathbf{x}+\mathbf{r})}.\mathbf{e}_z=\overline{\delta \rho(\mathbf{x}) \delta v(\mathbf{x}+\mathbf{r})}=B_{\rho,v}(\mathbf{r}),
\end{equation}
using the fact that $\overline{\mathbf{\delta v}(\mathbf{x})}=\mathbf{0}$. As a consequence, we have $B_{\rho,v}(-\mathbf{r})=-f(r)\mathbf{r}.\mathbf{e}_z=-B_{\rho,v}(\mathbf{r})$, so that $B_{\rho,v}$ is an antisymmetric field. Of course, this is a statistical property, and does not hold when considering the specific values of velocity and density fluctuations for a given pair of points within the flow. What one can conclude is that the integrand in Eq.~(\ref{eq_23}) is zero, and so $\left<y_{v_\mathbf{X}} y_{\rho_{\mathbf{X}+\mathbf{R}}} \right>+\left<y_{\rho_\mathbf{X}} y_{v_{\mathbf{X}+\mathbf{R}}} \right>=0$. So is the next term $\left<y_{\rho v_{\mathbf{X}\phantom{+}}}\!\!\!\right>$, since
\begin{equation}
\left<y_{\rho v_{\mathbf{X}\phantom{+}}}\!\!\!\right>=\frac{1}{\rho_0 v_0 D}\int\limits_{\!\!I_D} \! \left<\delta \rho(\mathbf{X},z) \delta v(\mathbf{X},z)\right> \mathrm{d} z=\frac{1}{\rho_0 v_0 D}\int\limits_{\!\!I_D} \! B_{\rho,v}(\mathbf{0}) \mathrm{d} z=0,
\end{equation}
because the antisymmetry of $B_{\rho,v}$ implies that $B_{\rho,v}(\mathbf{0})=0$. Similarly, the last two terms of $\left<a_1\right>$ are given by
\begin{equation}
\left<y_{\rho v_\mathbf{X}} y_{v_{\mathbf{X}+\mathbf{R}}} \right>+\left<y_{\rho v_{\mathbf{X}+\mathbf{R}}} y_{v_\mathbf{X}}\right>=\frac{1}{\rho_0v_0^2D^2}\int\!\!\!\!\int\limits_{\!\!I^2_D} \! \left[B_{\rho v,v}(\mathbf{R}+(z_2-z_1)\mathbf{e}_z)+B_{\rho v,v}(-\mathbf{R}+(z_1-z_2)\mathbf{e}_z)\right] \mathrm{d} z_1 \mathrm{d} z_2,
\end{equation}
introducing the two-point correlation function $B_{\rho v,v}(\mathbf{r})=\overline{\delta \rho(\mathbf{x})\delta v(\mathbf{x})\delta v(\mathbf{x}+\mathbf{r})}$. This last combination of terms, unlike the one considered previously, is not necessarily zero. According to~\citet{monin75}, it is of the form
\begin{equation}
B_{\rho v,v}(\mathbf{r})=g(r)\frac{(\mathbf{r}.\mathbf{e}_z)^2}{r^2}+h(r),
\end{equation}
where $g$ and $h$ are functions of the scalar separation $r$. This form stems from the fact that $\overline{\delta \rho(\mathbf{x})\delta v_i(\mathbf{x})\delta v_j(\mathbf{x}+\mathbf{r})}$, where $i$ and $j$ stand for any of the $x$, $y$ and $z$ coordinates, is a tensor of rank 2 which we suppose to be homogeneous and isotropic. It follows that $B_{\rho v,v}$ is symmetric, and so, using an averaging notation $M_{\rho v,v}$ defined, similarly to the expressions of $M_{\rho,\rho}$ and $M_{v,v}$, by
\begin{equation}
M_{\rho v,v} (\mathbf{R})=\frac{1}{D^2}\int\!\!\!\!\int\limits_{\!\!I^2_D} \! B_{\rho v,v}(\mathbf{R}+(z_2-z_1)\mathbf{e}_z)\mathrm{d} z_1 \mathrm{d} z_2,
\end{equation}
we can write $\left<a_1\right>$, contribution of the first order in density fluctuations to the autocorrelation function $A_C$, as
\begin{equation}
\left<a_1\right>=\left<y_{\rho v_\mathbf{X}} y_{v_{\mathbf{X}+\mathbf{R}}} \right>+\left<y_{\rho v_{\mathbf{X}+\mathbf{R}}} y_{v_\mathbf{X}}\right>=\frac{1}{\rho_0v_0^2}\left[M_{\rho v,v}(\mathbf{R})+M_{\rho v,v}(-\mathbf{R})\right]=\frac{2}{\rho_0v_0^2}M_{\rho v,v}(\mathbf{R}).
\end{equation}
Lastly, the second order term in density fluctuations $\left<a_2\right>=\left<y_{\rho_\mathbf{X}} y_{\rho_{\mathbf{X}+\mathbf{R}}}\right>+\left<y_{\rho v_\mathbf{X}} y_{\rho_{\mathbf{X}+\mathbf{R}}}\right>+\left<y_{\rho v_{\mathbf{X}+\mathbf{R}}}y_{\rho_\mathbf{X}}\right>+\left<y_{\rho v_\mathbf{X}} y_{\rho v_{\mathbf{X}+\mathbf{R}}}\right>$ is computed in much the same way, introducing the two-point correlation functions
\begin{equation}
B_{\rho v,\rho}(\mathbf{r})=\overline{\delta \rho(\mathbf{x})\delta v(\mathbf{x})\delta \rho(\mathbf{x}+\mathbf{r})} \quad \mathrm{and} \quad B_{\rho v,\rho v}(\mathbf{r})=\overline{\delta \rho(\mathbf{x})\delta v(\mathbf{x})\delta \rho(\mathbf{x}+\mathbf{r})\delta v(\mathbf{x}+\mathbf{r})},
\end{equation}
and their averages defined by
\begin{equation}
M_{\rho v,\rho} (\mathbf{R})=\frac{1}{D^2}\int\!\!\!\!\int\limits_{\!\!I^2_D} \! B_{\rho v,\rho}(\mathbf{R}+(z_2-z_1)\mathbf{e}_z)\mathrm{d} z_1 \mathrm{d} z_2 \quad \mathrm{and} \quad M_{\rho v,\rho v} (\mathbf{R})=\frac{1}{D^2}\int\!\!\!\!\int\limits_{\!\!I^2_D} \! B_{\rho v,\rho v}(\mathbf{R}+(z_2-z_1)\mathbf{e}_z)\mathrm{d} z_1 \mathrm{d} z_2.
\end{equation}
With these notations, it is straightforward to derive the expression of the second order term,
\begin{equation}
\left<a_2\right>=\frac{1}{\rho_0^2}\left[ M_{\rho,\rho}(\mathbf{R})+\frac{2}{v_0}M^{(s)}_{\rho v,\rho}(\mathbf{R})+\frac{1}{v^2_0} M_{\rho v,\rho v}(\mathbf{R})\right],
\end{equation}
with the $(s)$ superscript indicating the symmetric part of a given function, defined by the relation
\begin{equation}
f^{(s)}(x)=\frac{1}{2}\left[f(x)+f(-x)\right].
\end{equation}
Now, still following~\citet{monin75}, $B_{\rho v,\rho}$ exhibits the same antisymmetry property as $B_{\rho,v}$ and we can conclude that the symmetric part of $M_{\rho v,\rho}(\mathbf{R})$ is zero, so that the autocorrelation function $A_C$ of the non-normalized velocity centroid eventually reads 
\begin{equation}
\label{eq_003}
A_C(\mathbf{R})=(\alpha D)^2\left[\rho_0^2v_0^2+\rho_0^2M_{v,v}(\mathbf{R})+2\rho_0M_{\rho v,v}(\mathbf{R})+v_0^2M_{\rho,\rho}(\mathbf{R})+M_{\rho v,\rho v}(\mathbf{R})\right].
\end{equation}
When considering only the zeroth order in density fluctuations, one finds that the expression of $A_C(\mathbf{R})$ is very similar to that of $A_I(\mathbf{R})$, with $A_C(\mathbf{R}) \simeq (\alpha D\rho_0)^2\left[v_0^2+M_{v,v}(\mathbf{R})\right]$. This is not surprising, as we shall discuss in section~\ref{sec_test}. Moreover, as the forthcoming comparison of the autocorrelation functions of both normalized and non-normalized velocity centroids will be performed on expressions of order up to one only, it is useful to write out the truncation of $A_C(\mathbf{R})$ at that order, which is, given Eq.~(\ref{eq_003}),
\begin{equation}
\label{eq_003bis}
A_C(\mathbf{R})=(\alpha D)^2\left[\rho_0^2v_0^2+\rho_0^2M_{v,v}(\mathbf{R})+2\rho_0M_{\rho v,v}(\mathbf{R})\right].
\end{equation}

\subsection{Autocorrelation function of the normalized velocity centroid $C_0$}

There is little qualitative change to the method when dealing with the normalized centroid. Contributions of each order can be computed as easily as for the non-normalized case, with the main difference being that, given the expansion written in Eq.~(\ref{eq_201}), $A_{C_0}(\mathbf{R})$ is a theoretically infinite series which reads
\begin{equation}
A_{C_0}(\mathbf{R})=v_0^2\left<\left[\sum_{p,q}(-1)^{p+q}y^p_{\rho_\mathbf{X}}y^q_{\rho_{\mathbf{X}+\mathbf{R}}}\right]\left[\sum_{n=0}^2a_n\right]\right>=v_0^2\sum_{m}\sum_{n=0}^2\left<a_nb_m\right> \quad \mathrm{with} \quad b_m=(-1)^m\sum_{p=0}^m y^p_{\rho_\mathbf{X}}y^{m-p}_{\rho_{\mathbf{X}+\mathbf{R}}},
\end{equation}
the generic term $\left<a_nb_m\right>$ obviously being of order $n+m$ in density fluctuations. Now, given that the highest order present in $A_C(\mathbf{R})$ is two, it seems reasonable to consider only terms of order at most two in the expansion. However, it may also be argued that, since density fluctuations effectively contribute to the first order in the expression of $A_C(\mathbf{R})$, assessing the effects of normalization may already be performed when limiting the expansion to that order,
\begin{equation}
A_{C_0}(\mathbf{R})=v_0^2\left[\left<a_0b_0\right>+\left<a_0b_1\right>+\left<a_1b_0\right>\right]=v_0^2\left(S_0+S_1\right) \quad \mathrm{with} \quad S_0=\left<a_0b_0\right> \quad \mathrm{and} \quad S_1=\left<a_0b_1\right>+\left<a_1b_0\right>
\end{equation}
Now, the zeroth order term in $A_{C_0}(\mathbf{R})$ has been computed in the previous subsection,
\begin{equation}
S_0=\left<a_0b_0\right>=\left<a_0\right>=1+\frac{1}{v_0^2}M_{v,v}(\mathbf{R}),
\end{equation}
so we turn to the first order term $S_1=\left<a_0b_1\right>+\left<a_1b_0\right>$ which reads, after a long but straightforward calculation,
\begin{equation}
S_1=2\left<y_{\rho v_{\mathbf{X}\phantom{+}}}\!\!\!\right>-2\left<y_{\rho_{\mathbf{X}\phantom{+}}}\!\!\! y_{v_\mathbf{X}}\right>+\left<y_{\rho v_\mathbf{X}} y_{v_{\mathbf{X}+\mathbf{R}}} \right>+\left<y_{v_\mathbf{X}} y_{\rho v_{\mathbf{X}+\mathbf{R}}}\right>-\left<y_{v_\mathbf{X}} y_{v_{\mathbf{X}+\mathbf{R}}} y_{\rho_\mathbf{X}}\right>-\left<y_{v_\mathbf{X}} y_{v_{\mathbf{X}+\mathbf{R}}} y_{\rho_{\mathbf{X}+\mathbf{R}}}\right>.
\end{equation}
The first term is zero, as shown before. So is the second term, for we have
\begin{equation}
\left<y_{\rho_{\mathbf{X}\phantom{+}}}\!\!\! y_{v_\mathbf{X}}\right>=\frac{1}{\rho_0 v_0 D^2}\int\!\!\!\!\int\limits_{\!\!I^2_D} \! B_{\rho,v}((z_2-z_1)\mathbf{e}_z) \mathrm{d} z_1 \mathrm{d} z_2=0,
\end{equation}
since the integration domain is symmetric about $z_2-z_1=0$ and $B_{\rho,v}$ is antisymmetric. The combination of third and fourth terms has been computed in the previous subsection
\begin{equation}
\left<y_{\rho v_\mathbf{X}} y_{v_{\mathbf{X}+\mathbf{R}}} \right>+\left<y_{\rho v_{\mathbf{X}+\mathbf{R}}} y_{v_\mathbf{X}}\right>=\frac{2}{\rho_0v_0^2}M_{\rho v,v}(\mathbf{R}).
\end{equation}
The remaining terms require a more elaborate, although similar, treatment. Taking for instance the last term, we have
\begin{equation}
\left<y_{v_\mathbf{X}} y_{v_{\mathbf{X}+\mathbf{R}}} y_{\rho_{\mathbf{X}+\mathbf{R}}}\right>=\frac{1}{\rho_0v_0^2 D^3}\int\!\!\!\!\int\!\!\!\!\int\limits_{\!\!\!\!I^3_D} \! \left<\delta v(\mathbf{X},z_1) \delta v(\mathbf{X}+\mathbf{R},z_2)\delta \rho(\mathbf{X}+\mathbf{R},z_3)\right> \mathrm{d} z_1 \mathrm{d} z_2 \mathrm{d} z_3,
\end{equation} 
where the integration is performed over the cubic domain $I_D^3=I_D \times I_D \times I_D$. Introducing the three-point correlation function $B_{v,v,\rho}(\mathbf{r}_1,\mathbf{r}_2)=\overline{\delta v(\mathbf{x})\delta v(\mathbf{x}+\mathbf{r}_1)\delta \rho(\mathbf{x}+\mathbf{r}_2)}$, this expression reads
\begin{equation}
\left<y_{v_\mathbf{X}} y_{v_{\mathbf{X}+\mathbf{R}}} y_{\rho_{\mathbf{X}+\mathbf{R}}}\right>=\frac{1}{\rho_0v_0^2 D^3}\int\!\!\!\!\int\!\!\!\!\int\limits_{\!\!\!\!I^3_D} \!  B_{v,v,\rho}(\mathbf{R}+(z_2-z_1)\mathbf{e}_z,\mathbf{R}+(z_3-z_1)\mathbf{e}_z)\mathrm{d} z_1 \mathrm{d} z_2 \mathrm{d} z_3=\frac{1}{\rho_0v_0^2}M_{v,v,\rho}(\mathbf{R},\mathbf{R}),
\end{equation} 
with an averaging notation $M_{v,v,\rho}(\mathbf{R},\mathbf{R})$ reminiscent of the ones used previously, 
\begin{equation}
M_{v,v,\rho}(\mathbf{R},\mathbf{R})=\frac{1}{D^3}\int\!\!\!\!\int\!\!\!\!\int\limits_{\!\!\!\!I^3_D} \!  B_{v,v,\rho}(\mathbf{R}+(z_2-z_1)\mathbf{e}_z,\mathbf{R}+(z_3-z_1)\mathbf{e}_z)\mathrm{d} z_1 \mathrm{d} z_2 \mathrm{d} z_3.
\end{equation}
Similarly, the next-to-last term can be written as
\begin{equation}
\left<y_{v_\mathbf{X}} y_{v_{\mathbf{X}+\mathbf{R}}} y_{\rho_\mathbf{X}}\right>=\frac{1}{\rho_0v_0^2}M_{v,v,\rho}(-\mathbf{R},-\mathbf{R}) \quad \mathrm{so~that} \quad \left<y_{v_\mathbf{X}} y_{v_{\mathbf{X}+\mathbf{R}}} y_{\rho_{\mathbf{X}+\mathbf{R}}}\right>+\left<y_{v_\mathbf{X}} y_{v_{\mathbf{X}+\mathbf{R}}} y_{\rho_\mathbf{X}}\right>=\frac{2}{\rho_0v_0^2}M^{(s)}_{v,v,\rho}(\mathbf{R},\mathbf{R}),
\end{equation}
with $M^{(s)}_{v,v,\rho}(\mathbf{R},\mathbf{R})$ being the symmetric part of the function $M_{v,v,\rho}(\mathbf{R},\mathbf{R})$,
\begin{equation}
M^{(s)}_{v,v,\rho}(\mathbf{R},\mathbf{R})=\frac{1}{2}\left[M_{v,v,\rho}(\mathbf{R},\mathbf{R})+M_{v,v,\rho}(-\mathbf{R},-\mathbf{R})\right].
\end{equation}
Finally, the autocorrelation function $A_{C_0}(\mathbf{R})$ of the normalized centroid has the following expression when limited to contributions of the first order in density fluctuations,
\begin{equation}
\label{eq_004}
A_{C_0}(\mathbf{R})=v_0^2+M_{v,v}(\mathbf{R})+\frac{2}{\rho_0}\left[M_{\rho v,v}(\mathbf{R})-M^{(s)}_{v,v,\rho}(\mathbf{R},\mathbf{R})\right].
\end{equation}
It appears then that normalization of velocity centroids indeed performs a first order correction, as compared with Eq.~(\ref{eq_003bis}), although it does not necessarily fully remove the density structure, as will be discussed in more detail in section~\ref{sec_test}.

\section{The case of fractional Brownian motion fields}
\label{sec_fbm}

Explicitly computing the forms of the two-dimensional statistical measures $A_I(\mathbf{R})$, $A_C(\mathbf{R})$ and $A_{C_0}(\mathbf{R})$ requires the knowledge of the three-dimensional correlation functions appearing as integrated terms in equations~(\ref{eq_002}), (\ref{eq_003bis}) and (\ref{eq_004}). However, it may prove impossibly difficult to build a complete and consistent set of such functions and then to analytically compute their averages. Therefore, as a first step, one should stick to the lowest order terms in the expressions of the two-dimensional autocorrelation functions, 
\begin{equation}
\label{eq_001}
A_I(\mathbf{R})=(\alpha D)^2\left[\rho_0^2+M_{\rho,\rho}(\mathbf{R})\right], \qquad A_C(\mathbf{R})\simeq(\alpha D \rho_0)^2\left[v_0^2+M_{v,v}(\mathbf{R})\right] \qquad \mathrm{and} \qquad A_{C_0}(\mathbf{R}) \simeq v_0^2+M_{v,v}(\mathbf{R}).
\end{equation}
In this limit, one only has to compute the $M_{\rho,\rho}(\mathbf{R})$ and $M_{v,v}(\mathbf{R})$ averages given density and velocity fields with known autocorrelation functions $B_{\rho,\rho}(\mathbf{r})$ and $B_{v,v}(\mathbf{r})$. One possibility is to suppose that both of the three-dimensional fields are fractional Brownian motions (fBm). These are defined in the following way: if $\mathcal{F}$ is such a field in $N$ dimensions, it is characterized by a single-index power-law structure function,
\begin{equation}
\label{eq_101}
S_\mathcal{F}(\mathbf{r})=\overline{\left[\mathcal{F}(\mathbf{x}+\mathbf{r})-\mathcal{F}(\mathbf{x})\right]^2}=2\Lambda \left(\frac{r}{D}\right)^{2H},
\end{equation}
where $r=\left|\mathbf{r}\right|$ is the length of the separation vector, $\Lambda$ is a positive constant and $H$ is a real number in $[0,1]$ called the Hurst exponent. The restriction $H\geqslant 0$ corresponds to the fact that the amplitude of fluctuations should decrease at smaller scale, while we impose $H \leqslant 1$ because an fBm field having a Hurst exponent $H > 1$ would be uniform, as shown in appendix~\ref{sec_uniform}. Fractional Brownian motion fields have already been used~\citep{stutzki98,brunt2002a,mamd2003} to model clouds of the diffuse, non-starforming interstellar medium. In this section, we shall use them to model the three-dimensional density and velocity fields, so that, for the former,
\begin{equation}
S_\rho(\mathbf{r})=\overline{\left[\rho(\mathbf{x}+\mathbf{r})-\rho(\mathbf{x})\right]^2}=\overline{\left[\delta\rho(\mathbf{x}+\mathbf{r})-\delta\rho(\mathbf{x})\right]^2}=2\left[\sigma_\rho^2 -B_{\rho,\rho}(\mathbf{r})\right]=2\Lambda_\rho \left(\frac{r}{D}\right)^{2H_\rho},
\end{equation}
since the density fluctuations field is supposed to be homogeneous. Similarly, for the velocity field,
\begin{equation}
S_v(\mathbf{r})=\overline{\left[v(\mathbf{x}+\mathbf{r})-v(\mathbf{x})\right]^2}=\overline{\left[\delta v(\mathbf{x}+\mathbf{r})-\delta v(\mathbf{x})\right]^2}=2\left[\sigma_v^2 -B_{v,v}(\mathbf{r})\right]=2\Lambda_v \left(\frac{r}{D}\right)^{2H_v}.
\end{equation}
These last two equations make use of the relationship between the second order structure function $S_\mathcal{F}$ of an homogeneous field $\mathcal{F}$ and its autocorrelation function $A_\mathcal{F}$, which is $S_\mathcal{F}(\mathbf{R})=2\left[A_\mathcal{F}(\mathbf{0})-A_\mathcal{F}(\mathbf{R})\right]$. In the results of section~\ref{sec_norm}, we then have to inject the following relations
\begin{equation}
B_{\rho,\rho}(\mathbf{r})=\sigma_\rho^2-\Lambda_\rho \left(\frac{r}{D}\right)^{2H_\rho} \qquad \mathrm{and} \qquad B_{v,v}(\mathbf{r})=\sigma_v^2-\Lambda_v \left(\frac{r}{D}\right)^{2H_v}.
\end{equation}
To compute the averages $M_{v,v}(\mathbf{R})$ and $M_{\rho,\rho}(\mathbf{R})$ featured in the expressions of $A_I(\mathbf{R})$, $A_C(\mathbf{R})$ and $A_{C_0}(\mathbf{R})$, we can use the calculation scheme of~\citet{chandrasekhar52}, which is given in appendix~\ref{sec_chandra}, to write them as single integrals over the separation $\Delta z=z_2-z_1$ along the line of sight, noting that $\Delta z \in I_{2D}=[-D,D]$. We then have
\begin{equation}
M_{v,v}(\mathbf{R})=\frac{1}{D^2}\int\limits_{I_{2D}}\!\! (D-|\Delta z|) B_{v,v}(\mathbf{R}+\Delta z\mathbf{e}_z) \mathrm{d} \Delta z \quad \mathrm{and} \quad M_{\rho,\rho}(\mathbf{R})=\frac{1}{D^2}\int\limits_{I_{2D}}\!\! (D-|\Delta z|) B_{\rho,\rho}(\mathbf{R}+\Delta z\mathbf{e}_z) \mathrm{d} \Delta z.
\end{equation} 
In these expressions, the factor $(D-|\Delta z|)$ represents the weight of each separation $\Delta z$, that is the relative number of pairs of points taken into account whose projected separation along the line of sight is $\Delta z$. With the expressions above for $B_{v,v}$ and $B_{\rho,\rho}$, it is possible, as shown in appendix~\ref{sec_appendixfbm}, to write these averages under the form 
\begin{equation}
M_{v,v}(\mathbf{R})=-\Lambda_v K(R,H_v)+ \sigma_v^2 \quad \mathrm{and} \quad M_{\rho,\rho}(\mathbf{R})=-\Lambda_\rho K(R,H_\rho)+\sigma_\rho^2,
\end{equation}	
where $R=|\mathbf{R}|$ is the scalar separation on the plane of the sky, and the function $K$ is given by
\begin{equation}
\label{eq_207}
K(R,H)=\frac{2}{D^{2H+1}}\int_0^D\!\! \left(R^2+z^2\right)^{H} \mathrm{d} z-\frac{1}{H+1}\left[\left(1+\frac{R^2}{D^2}\right)^{H+1}\!\!\!\!\!\!-\left(\frac{R^2}{D^2}\right)^{H+1}\right].
\end{equation}
The integral in Eq.~(\ref{eq_207}) cannot be explicited~\citep{gradshteyn80} unless $H=0$ or $H=1$, but one interesting limit to consider is that of small separations on the plane of the sky ($R \ll D$), in which case $K(R,H)$ can be developed in powers of $R/D$. As shown in appendix~\ref{sec_appendixfbm}, this yields the following scaling relations for $M_{v,v}(\mathbf{R})$
\begin{equation}
M_{v,v}(\mathbf{R})\simeq \sigma_v^2-\Lambda_v a(H_v)-\Lambda_v b(H_v)\left(\frac{R}{D}\right)^{2H_v+1}\quad \mathrm{for} \quad 0 \leqslant H_v < \frac{1}{2},
\end{equation}
\begin{equation}
M_{v,v}(\mathbf{R})\simeq\sigma_v^2-\Lambda_v a(H_v)-\Lambda_v b(H_v)\left(\frac{R}{D}\right)^2\quad \mathrm{for} \quad \frac{1}{2} < H_v \leqslant 1, 
\end{equation}
\begin{equation}
M_{v,v}(\mathbf{R})\simeq\sigma_v^2-\frac{\Lambda_v }{3}+ \Lambda_v \left(\frac{R}{D}\right)^2\ln{\left(\frac{R}{D}\right)}\quad \mathrm{for} \quad H_v=\frac{1}{2},
\end{equation}
where $a$ and $b$ are functions of $H_v$ only. Similar relations are satisfied by $M_{\rho,\rho}(\mathbf{R})$ depending on the value of $H_\rho$. To the lowest order, the structure function $S_{C_0}(\mathbf{R})$ of the normalized centroid $C_0$ therefore reads
\begin{equation}
S_{C_0}(\mathbf{R})=2\Lambda_v b(H_v)\left(\frac{R}{D}\right)^{2H_v+1} \quad \mathrm{for} \quad 0 \leqslant H_v < \frac{1}{2},
\end{equation}
\begin{equation}
S_{C_0}(\mathbf{R})=2\Lambda_v b(H_v)\left(\frac{R}{D}\right)^{2}\quad \mathrm{for} \quad \frac{1}{2} < H_v \leqslant 1,
\end{equation}
\begin{equation}
S_{C_0}(\mathbf{R})=-2\Lambda_v \left(\frac{R}{D}\right)^2\ln{\left(\frac{R}{D}\right)}\quad \mathrm{for} \quad H_v=\frac{1}{2}.
\end{equation}
According to Eq.~(\ref{eq_001}), the structure function $S_C$ of the non-normalized velocity centroid has the same scaling behaviour. This is also the case for the structure function $S_I$ of the intensity, depending on the value of $H_\rho$. The consequences of such forms are presented in the next section.

\section{Discussion}
\label{sec_test}
As expected, the statistical measures on the intensity and centroid maps can be expressed in terms of the statistical properties of the underlying three-dimensional velocity and density fields. More precisely, it should come as no surprise that the autocorrelation functions $A_I(\mathbf{R})$, $A_C(\mathbf{R})$ and $A_{C_0}(\mathbf{R})$ should be written as averages of correlation functions within the slab in the manner described in Fig.~\ref{fig_2}, since all pairs of points with a given separation $\mathbf{R}$ in the plane of the sky should contribute to the two-dimensional statistical measures at lag $\mathbf{R}$. 
\\
It also appears that the zeroth order term in the autocorrelation functions of the velocity centroids has the same form as the autocorrelation of the intensity map, since we then have
\begin{equation}
A_{C}(\mathbf{R})\simeq (\alpha\rho_0 D)^2 A_{C_0}(\mathbf{R}) \simeq (\alpha\rho_0 D)^2\left[v_0^2+M_{v,v}(\mathbf{R})\right] \qquad \mathrm{while} \qquad A_I(\mathbf{R})=(\alpha D)^2\left[\rho_0^2+M_{\rho,\rho}(\mathbf{R})\right].
\end{equation}
The identity of these forms was to be expected, since the zeroth order terms in the autocorrelation functions of the velocity centroids are the limits obtained when the density is uniform. In this case, the centroids simply are integrals of $v$, as the intensity is a simple integral of $\rho$. In the context of fBm fields, the implication of this limit is that both non-normalized and normalized velocity centroids have a fractional Brownian motion behaviour, since their structure functions are power laws, with respective Hurst exponents $H_C$ and $H_{C_0}$ such that
\begin{equation}
H_C=H_{C_0}=H_v+\frac{1}{2}\quad \mathrm{for} \quad 0 \leqslant H_v < \frac{1}{2} \qquad \mathrm{and} \qquad 
H_C=H_{C_0}=1 \quad \mathrm{for} \quad \frac{1}{2} < H_v \leqslant 1.
\end{equation}
And similarly, the intensity map has a fractional Brownian motion behaviour with a Hurst exponent $H_I$ with
\begin{equation}
H_I=H_\rho+\frac{1}{2}\quad \mathrm{for} \quad 0 \leqslant H_\rho < \frac{1}{2} \qquad \mathrm{and} \qquad 
H_I=1 \quad \mathrm{for} \quad \frac{1}{2} < H_\rho \leqslant 1.
\end{equation}
These results should be interpreted in the light of what is already known about the statistical properties of the integrated emission map $I$ from studies in the Fourier domain. Indeed, in the case of optically thin lines, the power spectrum index $\gamma_I$ of the intensity map is the same as that of the three-dimensional density field $\gamma_\rho$~\citep[see e.g.][]{goldman2000} provided the depth probed $D$ is larger than the transverse scales $R$, which is the case in the limit considered in section~\ref{sec_fbm}. Now, spectral index $\gamma$ and Hurst exponent $H$ are related by $\gamma=2H+N$, with $N$ the dimension of the space over which the field is defined. As a result,~\citet{stutzki98} concluded that the Hurst exponent of the integrated emission map is $H_I=H_\rho+1/2$ for $H_\rho \leqslant 1/2$ with a turnover to $H_I=1$ for $H_\rho \geqslant 1/2$, since
\begin{equation}
H_I=\frac{\gamma_I-2}{2}=\frac{\gamma_\rho-3}{2}+\frac{1}{2}=H_\rho+\frac{1}{2},
\end{equation}
using the fact that the integrated emission map is two-dimensional, while the original density field is three-dimensional. This result is precisely what we find from our analytical study performed solely in real space. Similarly, the expression for the structure function of the normalized centroid confirms the numerical findings of~\citet{mamd2003}, who showed that the spectral index $\gamma_{C_0}$ of normalized centroid maps was equal to that of the three-dimensional velocity field, $\gamma_v$. It should be noted that their result holds even for density fields with large fluctuations, which implies that our analytical result may be applicable to a more general class of fields. On the other hand, in the limit of negligible density fluctuations, one should find the same behaviour for the non-normalized centroid, a result~\citet{mamd2003} have not reported on.
 
Returning to a study of the autocorrelation functions of velocity centroids in the first order approximation,
\begin{displaymath}
A_C(\mathbf{R})\simeq(\alpha D\rho_0)^2\left[v_0^2+M_{v,v}(\mathbf{R})+\frac{2}{\rho_0}M_{\rho v,v}(\mathbf{R})\right] \quad \mathrm{and} \quad A_{C_0}(\mathbf{R})\simeq v_0^2+M_{v,v}(\mathbf{R})+\frac{2}{\rho_0}\left[M_{\rho v,v}(\mathbf{R})-M^{(s)}_{v,v,\rho}(\mathbf{R},\mathbf{R})\right],
\end{displaymath}
we see that, since $M_{\rho v,v}(\mathbf{R})-M^{(s)}_{v,v,\rho}(\mathbf{R},\mathbf{R})$ is not necessarily zero, normalization actually does not remove the first order contribution of density fluctuations. The empirical assumption that normalization somehow eliminates the influence of density fluctuations on velocity centroids may therefore not be true, although assessing the magnitude of the remaining first order contribution in the normalized centroid's autocorrelation function with respect to the non-normalized case could be difficult, as it involves computing averages of correlation functions whose forms are still unknown. This will be investigated in a forthcoming paper.

Such shortcomings of the normalized centroids as a way to retrieve the actual velocity statistics from the observational data have already been pointed out by~\citet{lazarian2003}, albeit in the case of simulations of highly compressible MHD turbulence. Instead, they introduced modified velocity centroids (MVC) $C_m$, defined through their second order structure function $S_{C_m}$ as
\begin{equation}
\label{eq_0009}
S_{C_m}(\mathbf{R})=\left<\left[C_m(\mathbf{X}+\mathbf{R})-C_m(\mathbf{X})\right]^2\right>=S_C(\mathbf{R})-(v_0^2+\sigma_v^2+\sigma_\mathrm{th}^2)S_I(\mathbf{R}),
\end{equation}
where $S_C$ and $S_I$ are the structure functions of the non-normalized velocity centroid and of the intensity, respectively. Given the relationship between the structure and autocorrelation functions, we can write the autocorrelation function $A_{C_m}$ of the modified velocity centroids as
\begin{equation}
A_{C_m}(\mathbf{R})=A_C(\mathbf{R})-(v_0^2+\sigma_v^2+\sigma_\mathrm{th}^2)A_I(\mathbf{R})+E,
\end{equation}
$E$ being a constant involving the values of $A_C$ and $A_I$ at zero lag. 
Now, since $A_I$ contains no first-order term, it appears that modified and non-normalized velocity centroids have the same first order contribution in their autocorrelation functions. As noted earlier, however, it is not yet clear whether this contribution is actually more important than the first order term in the autocorrelation function of normalized centroids. If it is, it would tend to prove that normalization may be a better way of retrieving velocity statistics in weakly compressible flows. To make the comparison of correction schemes clearer, one can show a relationship between the autocorrelation functions of $I$, $C$ and $C_0$ as written in equations~(\ref{eq_002}), (\ref{eq_003}) and (\ref{eq_004}),
\begin{equation}
\label{eq_205}
(\alpha D \rho_0)^2 A_{C_0}(\mathbf{R})=(\alpha D \rho_0v_0)^2+A_C(\mathbf{R})-v_0^2A_I(\mathbf{R})-2(\alpha D)^2\rho_0M^{(s)}_{v,v,\rho}(\mathbf{R},\mathbf{R})+G(\mathbf{R}),
\end{equation}
where $G(\mathbf{R})$ stands for the terms of order at least two in density fluctuations. Deriving the relationship between the structure functions of $I$, $C$ and $C_0$, the latter being noted $S_{C_0}$, we have
\begin{equation}
(\alpha D \rho_0)^2 S_{C_0}(\mathbf{R})=S_C(\mathbf{R})-v_0^2S_I(\mathbf{R})-4(\alpha D)^2\rho_0\left[M^{(s)}_{v,v,\rho}(\mathbf{0},\mathbf{0})-M^{(s)}_{v,v,\rho}(\mathbf{R},\mathbf{R})\right]+2\left[G(\mathbf{0})-G(\mathbf{R})\right].
\end{equation}
Comparing this with Eq.~(\ref{eq_0009}), it is clear that modified velocity centroids subtract more of the intensity structure, thus removing second order terms, while normalization performs a more complex correction, involving all orders in density fluctuations. Actual comparison of the merits of normalized and modified velocity centroids requires numerical tests, which have only been performed by~\citet{lazarian2003} on highly compressible MHD turbulence. That particular case is actually not within the scope of our study, firstly because density fluctuations are large and the condition for the expansion in the normalized velocity centroid's case may not be met, and secondly because the presence of a magnetic field implies anisotropy. Both limitations render the comparison with the work of~\citet{lazarian2003} a bit uncertain, although it is likely that terms of higher order in density fluctuations may become dominant. Thus, the effectiveness of modified velocity centroids, as compared to normalized centroids, in their simulations may be related to the second order correction through $S_I$, which is more important in modified velocity centroids. In any case, further analytical and numerical work is warranted in order to establish these conclusions more firmly, especially in the weakly compressible flow regime.

\section{Conclusions}
\label{sec_concl}

An analytical study of a simple slightly compressible turbulent cloud model was presented, assuming homogeneity and isotropy of the turbulent flow. From the expressions of the antenna temperature for an optically thin spectral line and of its successive moments with respect to the line of sight velocity component, we computed the autocorrelation functions of the intensity and of both normalized and non-normalized velocity centroids, which involve averages, along the line of sight, of correlation functions of the three-dimensional density and velocity fields. 

To the lowest order, the autocorrelation functions of the velocity centroids behave, with respect to the velocity field, as the autocorrelation function of the intensity with respect to the density field. This sheds light on the numerical result of~\citet{mamd2003}, who found that, for fractional Brownian motion density and velocity fields, the spectral index of the normalized centroid is equal to that of the velocity field. We derived this result analytically, for separations across the sky much smaller than the cloud's depth, and in real space, while previous studies such as that of~\citet{goldman2000} were performed in Fourier space. However, the result of~\citet{mamd2003} holds for fields outside of the validity domain for our calculation.

Comparison of the expansions of the autocorrelation functions of both types of velocity centroids shows that normalization performs a correction of the first order in density fluctuations, although its magnitude remains to be assessed, a task for which numerical simulations are probably necessary. Numerical tests should also provide us with a robust comparison between normalized velocity centroids and the modified velocity centroids of~\citet{lazarian2003}, which imply corrections of order two in density fluctuations. At present, this comparison has been performed only on simulations of highly compressible and magnetized turbulence, a case beyond the scope of our analytical study, and has shown that, in this particular case, modified velocity centroids provide a more reliable tool than normalized centroids.

In a forthcoming paper, we shall therefore present numerical simulations aimed at assessing the validity domain of our calculations and, beyond normalized and modified velocity centroids, pursuing the search for a better correction scheme able to retrieve the underlying velocity statistics from observational data.  
\\
%\begin{acknowledgements}
I wish to acknowledge fruitful discussions with Alex Lazarian during his stay at the \'Ecole normale sup\'erieure. His suggestions, in the early stages of this work, proved very helpful. I also wish to thank Enrique V\'azquez-Semadeni for his careful reading of the manuscript and insightful remarks.
%\end{acknowledgements}

\appendix

%\onecolumn

\section{The Chandrasekhar-M\"unch scheme}
\label{sec_chandra}

For a function $f(\mathbf{r})$ depending on the three-dimensional separation $\mathbf{r}$, double integrals of the form
\begin{equation}
\mathcal{I}=\int\!\!\!\!\int\limits_{\!\!I^2_D} \! f(\mathbf{R}+(z_2-z_1)\mathbf{e}_z)\mathrm{d} z_1 \mathrm{d} z_2
\end{equation}
can be transformed, taking into account the fact that the integrand only depends on the difference $\Delta z=z_2-z_1$, because $\mathbf{R}$ is fixed, as shown in~\citet{chandrasekhar52}. Making the variable change $\Delta z=z_2-z_1$ and $z=z_1$,
\begin{equation}
\mathcal{I}=\int\limits_{I_D} \!\! \mathrm{d} z \!\! \int\limits_{I_{D_z}} \!\! \mathrm{d} \Delta z\, f(\mathbf{R}+\Delta z\mathbf{e}_z).
\end{equation}
where the notation $I_{D_z}$ stands for the segment $I_D$ shifted by $-z$, $I_{D_z}=[-D/2-z,D/2-z]$, which we can separate in two, with a negative part, $I_{D_z}^-=[-D/2-z,0]$, and a positive part $I_{D_z}^+=[0,D/2-z]$.
\begin{equation}
\mathcal{I}=\int\limits_{I_D} \!\! \mathrm{d} z \!\!  \int\limits_{I_{D_z}^-}\!\! \mathrm{d} \Delta z\, f(\mathbf{R}+\Delta z\mathbf{e}_z)+\int\limits_{I_D} \!\! \mathrm{d} z \!\!  \int\limits_{I_{D_z}^-}\!\! \mathrm{d} \Delta z\, f(\mathbf{R}+\Delta z\mathbf{e}_z).
\end{equation}
Exchanging the order of integrations in each of these two terms, and expliciting the integral over $z$, we have
\begin{equation}
\mathcal{I}=\int_{-D}^0\!\!\!\!\!\! f(\mathbf{R}+\Delta z\mathbf{e}_z)(D+\Delta z) \mathrm{d} \Delta z+\int_{0}^D\!\!\!\!\!\! f(\mathbf{R}+\Delta z\mathbf{e}_z)(D-\Delta z) \mathrm{d} \Delta z,
\end{equation}
which can be finally written as
\begin{equation}
\mathcal{I}=\int\limits_{I_{2D}}\!\! (D-|\Delta z|) f(\mathbf{R}+\Delta z\mathbf{e}_z) \mathrm{d} \Delta z.
\end{equation}
The factor $(D-|\Delta z|)$ expresses the fact that for a finite slab, two given separations $\Delta z$ and $\Delta z'$ are not represented by the same number of pairs of points.

\section{Autocorrelation functions of the velocity centroid maps in the case $v_0=0$}
\label{sec_zero}

When the mean velocity $v_0$ is null, the method used in the main body of the paper cannot be applied without some modification. Little needs to be done, however, as velocity fluctuations can be measured with respect to the sound speed $c_s$, which we assume to be uniform within the turbulent slab. We may then write the velocity centroid $C$ as
\begin{equation}
C(\mathbf{X})=\alpha\!\!\int\limits_{I_D} \!\!\! \left(\rho_0+\delta \rho(\mathbf{X},z)\right)  \delta v(\mathbf{X},z) \mathrm{d} z = \alpha \rho_0 c_s D \left[ \frac{1}{D}\int\limits_{I_D} \!\frac{\delta v(\mathbf{X},z)}{c_s}\mathrm{d} z + \frac{1}{D}\int\limits_{I_D} \!\frac{\delta \rho(\mathbf{X},z)}{\rho_0}\frac{\delta v(\mathbf{X},z)}{c_s}\mathrm{d} z\right],
\end{equation}
which can be written in the shorthand form $C(\mathbf{X})=\alpha \rho_0 c_s D\left[y_{v_\mathbf{X}}+y_{\rho v_\mathbf{X}}\right]$. The normalized velocity centroid then reads
\begin{equation}
C_0(\mathbf{X})=\frac{C(\mathbf{X})}{I(\mathbf{X})}=c_s\left[y_{v_\mathbf{X}}+y_{\rho v_\mathbf{X}}\right]\left[1+y_{\rho_\mathbf{X}}\right]^{-1}.
\end{equation}
The calculation therefore proceeds in the same way, with $c_s$ taking the place of $v_0$ and the $a_i$ of Eq.~(\ref{eq_11}) being replaced by new coefficients $a'_i$ given by
\begin{displaymath}
\begin{array}{l}
a'_0= y_{v_\mathbf{X}} y_{v_{\mathbf{X}+\mathbf{R}}} , \\
a'_1= y_{\rho v_\mathbf{X}} y_{v_{\mathbf{X}+\mathbf{R}}} +y_{\rho v_{\mathbf{X}+\mathbf{R}}} y_{v_\mathbf{X}}, \\
a'_2=y_{\rho v_\mathbf{X}} y_{\rho v_{\mathbf{X}+\mathbf{R}}}.
\end{array}
\end{displaymath}
It follows that the autocorrelation function of the non-normalized centroid is, in this case,
\begin{equation}
A_C(\mathbf{R})=\left(\alpha D\right)^2\left[\rho_0^2M_{v,v}(\mathbf{R})+2\rho_0M_{\rho v,v}(\mathbf{R})+M_{\rho v,\rho v}(\mathbf{R})\right], 
\end{equation}
which is precisely what is found form the general case when setting $v_0=0$. Similarly, for the normalized centroid,
\begin{equation}
A_{C_0}(\mathbf{R})=M_{v,v}(\mathbf{R})+\frac{2}{\rho_0}\left[M_{\rho v,v}(\mathbf{R})-M^{(s)}_{v,v,\rho}(\mathbf{R},\mathbf{R})\right].
\end{equation}

\section{Expansion of the averaged autocorrelation function of an fBm field}
\label{sec_appendixfbm}

We wish to compute the following expression as a function of the lag $R=\left|\mathbf{R}\right|$, in the limit $R \ll D$,
\begin{equation}
M(\mathbf{R})=\frac{1}{D^2}\int\limits_{I_{2D}}\left[-\Lambda \frac{\left(R^2+z^2\right)^{H}}{D^{2H}} + \sigma^2\right](D-|z|) \mathrm{d} z,
\end{equation}
where $\Lambda$ is a positive constant and $H \in [0,1]$ is the Hurst exponent. $M(\mathbf{R})$ can be expressed as an integral over $[0,D]$,
\begin{displaymath}
M(\mathbf{R})=\frac{2}{D^2}\int_0^D\left[-\Lambda \frac{\left(R^2+z^2\right)^{H}}{D^{2H}} + \sigma^2\right](D-z) \mathrm{d} z=\frac{2}{D^2}\left[-\Lambda \int_0^D\!\! \frac{\left(R^2+z^2\right)^{H}}{D^{2H}}(D-z) \mathrm{d} z + \sigma^2 \int_0^D\!\! (D-z) \mathrm{d} z \right].
\end{displaymath}
The last integral is easily shown to be equal to $D^2/2$, and so
\begin{equation}
M(\mathbf{R})=-\frac{2\Lambda}{D^{2H+1}} \int_0^D\!\! \left(R^2+z^2\right)^{H} \mathrm{d} z+\frac{2\Lambda}{D^{2H+2}} \int_0^D\!\! \left(R^2+z^2\right)^{H}z \mathrm{d} z + \sigma^2.
\end{equation}
The second of the remaining integrals can be explicited by setting $R^2+z^2=r^2$, since $z\mathrm{d} z=r\mathrm{d} r$ for $R$ fixed,
\begin{equation}
\int_0^D\!\! \left(R^2+z^2\right)^{H}z \mathrm{d} z=\int_R^{\sqrt{R^2+D^2}}\!\!\!\!\!\!\!\!\!\!\!\!\!\!\!\!\!\! r^{2H+1} \mathrm{d} r=\frac{1}{2H+2}\left[\left(R^2+D^2\right)^{H+1}-\left(R^2\right)^{H+1}\right],
\end{equation}
since $2H+1 \neq -1$ for $0 \leqslant H \leqslant 1$. We therefore have
\begin{equation}
\label{eq_50}
M(\mathbf{R})=-\frac{2\Lambda}{D^{2H+1}} \int_0^D\!\! \left(R^2+z^2\right)^{H} \mathrm{d} z+\frac{\Lambda}{H+1}\left[\left(1+\frac{R^2}{D^2}\right)^{H+1}-\left(\frac{R^2}{D^2}\right)^{H+1}\right] + \sigma^2.
\end{equation}
For a zero separation in the plane of the sky, this expression is, explicitly,
\begin{equation}
M(\mathbf{0})=-\frac{2\Lambda}{D^{2H+1}} \int_0^D\!\! z^{2H} \mathrm{d} z+\frac{\Lambda}{H+1} + \sigma^2=-\frac{\Lambda}{(2H+1)(H+1)} + \sigma^2.
\end{equation}
When $R \neq 0$, the first integral in Eq.~(\ref{eq_50}) generally cannot be written in closed form~\citep{gradshteyn80}. Notable exceptions are for $H=0$ and $H=1$, when, respectively,
\begin{equation}
M(\mathbf{R})=-\Lambda+ \sigma^2 \qquad \mathrm{and} \qquad M(\mathbf{R})=-\Lambda \left(\frac{R}{D}\right)^2-\frac{\Lambda}{6}+ \sigma^2.
\end{equation}
To simplify the notations, we introduce the function $K=K'-K''$, defined by $M(\mathbf{R})=-\Lambda K(R,H)+ \sigma^2$, with
\begin{equation}
K'(R,H)=\frac{2}{D^{2H+1}}\int_0^D\!\! \left(R^2+z^2\right)^{H} \mathrm{d} z \quad \mathrm{and} \quad K''(R,H)=\frac{1}{H+1}\left[\left(1+\frac{R^2}{D^2}\right)^{H+1}-\left(\frac{R^2}{D^2}\right)^{H+1}\right].
\end{equation}
It is now interesting to consider the case of separations $R \ll D$, which corresponds to studying the small scale structure of centroid maps. This allows to develop $K(R,H)$ in powers of $R/D$. Expansion of $K''$ is straightforward,
\begin{equation}
K''(R,H)=\frac{1}{H+1}\sum_{n\geqslant 0}\frac{\gamma_n(H+1)}{n!}\left(\frac{R}{D}\right)^{2n} \quad \mathrm{for} \quad 0 < H \leqslant 1 \qquad \mathrm{and} \qquad K''(R,0)=1,
\end{equation}
where we introduced $\gamma_n(x)=x(x-1)\ldots(x-n+1)$. As for $K'$, since we consider $D \gg R > 0$ we can write it as
\begin{equation}
K'(R,H)=2\int_0^{D/R}\!\! \left(1+y^2\right)^{H} \mathrm{d} y\left(\frac{R}{D}\right)^{2H+1}=2\left(\frac{R}{D}\right)^{2H+1}\left[\int_0^1\!\! \left(1+y^2\right)^{H} \mathrm{d} y+\int_1^{D/R}\!\! \left(1+y^2\right)^{H} \mathrm{d} y\right].
\end{equation}
The first integral, between $0$ and $1$, which we dub $K_0(H)$, cannot be explicited, but since it does not depend on $R$, it is unimportant with respect to the structure of the moment maps. The second integral can be transformed in order to develop the integrand through $1+y^2=y^2(1+y^{-2})$,
\begin{equation}
K'(R,H)=2\left(\frac{R}{D}\right)^{2H+1}\left[K_0(H)+ \sum_{k \geqslant 0} \frac{\gamma_k(H)}{k!} \int_1^{D/R}\!\! y^{2H-2k} \mathrm{d} y\right].
\end{equation}
Computing the integrals in the equation above for $0 < H < 1$ and  $H \neq 0.5$, we have
\begin{equation}
K'(R,H)=2\left(\frac{R}{D}\right)^{2H+1}\left\{K_0(H)+ \sum_{k \geqslant 0}\frac{\gamma_k(H)}{k!} \frac{1}{2H-2k+1}\left[\left(\frac{D}{R}\right)^{2H-2k+1}-1\right]\right\},
\end{equation}
which, after rearranging the terms, leads to the following expression for $K'$,
\begin{equation}
K'(R,H)=2\left[K_0(H)- \sum_{k \geqslant 0}\frac{\gamma_k(H)}{(2H-2k+1)k!}\right] \left(\frac{R}{D}\right)^{2H+1}+ \sum_{k \geqslant 0}\frac{2\gamma_k(H)}{(2H-2k+1)k!}\left(\frac{R}{D}\right)^{2k}.
\end{equation}
If $0 < H < 0.5$, then $1 < 2H+1 < 2$ and the leading order expansion is therefore
\begin{equation}
K'(R,H)\simeq\frac{2}{2H+1} + 2\left[K_0(H)-\sum_{k \geqslant 0}\frac{\gamma_k(H)}{2H-2k+1}\right] \left(\frac{R}{D}\right)^{2H+1}, 
\end{equation}
while if $0.5 < H < 1$, then $2 < 2H+1 < 3$ and the $k=1$ term of the last sum is dominant,
\begin{equation}
K'(R,H)\simeq\frac{2}{2H+1} + \frac{2H}{2H-1}\left(\frac{R}{D}\right)^2.
\end{equation}
This last expression is also valid for $H=1$, while obviously $K'(R,0)=2$. For $H=0.5$, the $k=1$ integral is
\begin{equation}
\int_1^{D/R}\!\! y^{-1} \mathrm{d} y=-\ln{\left(\frac{R}{D}\right)},
\end{equation}
and for $k \neq 1$ the integrals are unchanged. It follows that
\begin{equation}
K'(R,0.5)=2\left(\frac{R}{D}\right)^2\left\{K_0(0.5)- \frac{1}{2}\ln{\left(\frac{R}{D}\right)} +\sum_{k \neq 1}\frac{\gamma_k(0.5)}{2-2k}\left[\left(\frac{D}{R}\right)^{2-2k}\!\!\!\!\!\!-1\right]\right\},
\end{equation}
which gives, after rearranging the terms,
\begin{equation}
K'(R,0.5)=1- \left(\frac{R}{D}\right)^2\ln{\left(\frac{R}{D}\right)} +\left[2K_0(0.5)-\sum_{k \neq 1}\frac{\gamma_k(0.5)}{1-k}\right]\frac{R^2}{D^2}+\sum_{k \geqslant 2}\frac{\gamma_k(0.5)}{1-k}\left(\frac{R}{D}\right)^{2k}.
\end{equation}
Finally, the leading order expansion of $K(R,H)$ for small separations $R \ll D$ is
\begin{equation}
K(R,H)\simeq\frac{1}{(2H+1)(H+1)} + 2\left[K_0(H)- \sum_{k \geqslant 0}\frac{\gamma_k(H)}{2H-2k+1}\right] \left(\frac{R}{D}\right)^{2H+1} \quad \mathrm{for} \quad 0 \leqslant H < 0.5,
\end{equation}
\begin{equation}
K(R,H)\simeq\frac{1}{(2H+1)(H+1)} + \frac{1}{2H-1}\left(\frac{R}{D}\right)^2\quad \mathrm{for} \quad 0.5 < H \leqslant 1, 
\end{equation}
\begin{equation}
K(R,0.5)\simeq\frac{1}{3}- \left(\frac{R}{D}\right)^2\ln{\left(\frac{R}{D}\right)}.
\end{equation}
Consequently, the expansion of $M(\mathbf{R})=-\Lambda K(R,H)+ \sigma^2$ reads
\begin{equation}
M(\mathbf{R})\simeq \sigma^2-\frac{\Lambda}{(2H+1)(H+1)} -2\Lambda\left[K_0(H)- \sum_{k \geqslant 0}\frac{\gamma_k(H)}{2H-2k+1}\right] \left(\frac{R}{D}\right)^{2H+1} \quad \mathrm{for} \quad 0 \leqslant H < 0.5,
\end{equation}
\begin{equation}
M(\mathbf{R})\simeq\sigma^2-\frac{\Lambda}{(2H+1)(H+1)} - \frac{\Lambda}{2H-1}\left(\frac{R}{D}\right)^2\quad \mathrm{for} \quad 0.5 < H \leqslant 1, 
\end{equation}
\begin{equation}
M(\mathbf{R})\simeq\sigma^2-\frac{\Lambda}{3}+ \Lambda \left(\frac{R}{D}\right)^2\ln{\left(\frac{R}{D}\right)}\quad \mathrm{for} \quad H=0.5.
\end{equation}
The consequences on the statistical measures performed on the moment maps are straightforward to derive. They are given and analyzed in the main body of the paper.

\section{The uniformity of fBm fields with $H > 1$}
\label{sec_uniform}

We consider the mean squared increment of an fBm field $\mathcal{F}$ of Hurst exponent $H > 1$ between positions $\mathbf{x}$ and $\mathbf{x}+\mathbf{r}$. The separation vector $\mathbf{r}$ is then separated in $p$ equal parts, so that
\begin{equation}
\overline{\left[\mathcal{F}(\mathbf{x}+\mathbf{r})-\mathcal{F}(\mathbf{x})\right]^2}=\overline{\left\{\sum_{k=0}^{p-1}\left[\mathcal{F}\left(\mathbf{x}+\frac{k+1}{p}\mathbf{r}\right)-\mathcal{F}\left(\mathbf{x}+\frac{k}{p}\mathbf{r}\right)\right]\right\}^2}=\overline{\left[\sum_{k=0}^{p-1}\Delta\mathcal{F}_k\right]^2},
\end{equation}
where $\Delta\mathcal{F}_k$ is the increment of $\mathcal{F}$ between the $k^{\mathrm{th}}$ and $(k+1)^{\mathrm{th}}$ positions. Expansion of the expression above yields
\begin{equation}
\overline{\left[\mathcal{F}(\mathbf{x}+\mathbf{r})-\mathcal{F}(\mathbf{x})\right]^2}=\sum_{k=0}^{p-1}\overline{\Delta\mathcal{F}_k^2}+2\sum_{i<j}\overline{\Delta\mathcal{F}_i\Delta\mathcal{F}_j}.
\end{equation}
Now, $\overline{\Delta\mathcal{F}_i\Delta\mathcal{F}_j}$ can be written as an autocorrelation product of the function $\mathcal{G}_{\mathbf{r},p}=\mathcal{F}\left(\mathbf{x}+\mathbf{r}/p\right)-\mathcal{F}\left(\mathbf{x}\right)$,
\begin{equation}
\overline{\Delta\mathcal{F}_i\Delta\mathcal{F}_j}=\overline{\left[\mathcal{G}_{\mathbf{r},p}\left(\mathbf{x}+\frac{i}{p}\mathbf{r}\right)\right]\left[\mathcal{G}_{\mathbf{r},p}\left(\mathbf{x}+\frac{j}{p}\mathbf{r}\right)\right]}.
\end{equation}
Since autocorrelation functions are decreasing from their zero spacing value, we have
\begin{equation}
\overline{\Delta\mathcal{F}_i\Delta\mathcal{F}_j}\leqslant\overline{\left[\mathcal{G}_{\mathbf{r},p}\left(\mathbf{x}\right)\right]^2},
\end{equation}
which allows us to give an upper limit for the mean squared increment of $\mathcal{F}$,
\begin{equation}
\overline{\left[\mathcal{F}(\mathbf{x}+\mathbf{r})-\mathcal{F}(\mathbf{x})\right]^2} \leqslant \frac{2\Lambda}{p^{2H-1}}r^{2H}+p(p-1)\frac{2\Lambda}{p^{2H}}r^{2H}=2\Lambda p^{2(1-H)}r^{2H}.
\end{equation}
This result being valid for any value of $p$, which characterizes a subdivision of the separation vector $\mathbf{r}$, we see that since $H > 1$, the limit $p \to \infty$ implies $\overline{\left[\mathcal{F}(\mathbf{x}+\mathbf{r})-\mathcal{F}(\mathbf{x})\right]^2}=0$ and therefore $\mathcal{F}$ is a uniform field.

\end{document}